\documentclass[11pt,letterpaper]{article}
\pdfoutput=1
\usepackage{jcappub}
\usepackage{color}
\usepackage{graphicx}
\usepackage{wrapfig}

\usepackage{cancel}

\usepackage{verbatim}
\usepackage{amsmath} 
\usepackage{amssymb}
\usepackage{subfigure}
\usepackage{url}
\usepackage{bbold}
\usepackage{xspace}

\usepackage{multirow}
\usepackage{threeparttable}
\usepackage{paralist}

\newcommand{\GeV}{\text{GeV}}
\newcommand{\TeV}{\text{TeV}}

\DeclareRobustCommand{\Sec}[1]{Sec.~\ref{#1}}

\DeclareRobustCommand{\App}[1]{App.~\ref{#1}}

\DeclareRobustCommand{\Fig}[1]{Fig.~\ref{#1}}

\DeclareRobustCommand{\Eq}[1]{Eq.~(\ref{#1})}
\DeclareRobustCommand{\Eqs}[2]{Eqs.~(\ref{#1}) and (\ref{#2})}

\newcommand{\be}{\begin{equation}}
\newcommand{\ee}{\end{equation}}
\newcommand{\bea}{\begin{eqnarray}}
\newcommand{\eea}{\end{eqnarray}}

\newcommand{\mb}[1]{\boldsymbol{#1}}

\newcommand{\vmin}{v_{min}}

\def\ltap{\ \raise.3ex\hbox{$<$\kern-.75em\lower1ex\hbox{$\sim$}}\ }
\def\gtap{\ \raise.3ex\hbox{$>$\kern-.75em\lower1ex\hbox{$\sim$}}\ }
\def\lsim{\ \raise.3ex\hbox{$<$\kern-.75em\lower1ex\hbox{$\sim$}}\ }
\def\gsim{\ \raise.3ex\hbox{$>$\kern-.75em\lower1ex\hbox{$\sim$}}\ }
\def\eg{{\it e.g.}}
\def\ie{{\it i.e.}}

\usepackage[usenames,dvipsnames]{xcolor}

\begin{document}
\begin{flushright}
CERN-PH-TH-2015-073, FERMILAB-PUB-15-096-T, MIT-CTP/4661
\end{flushright}

\title{Halo-Independent Direct Detection Analyses Without Mass Assumptions}

\author[a]{Adam J. Anderson,}

\affiliation[a]{Laboratory for Nuclear Science, Massachusetts Institute of Technology,\\Cambridge, MA 02139, USA}

\author[b]{Patrick J. Fox,}

\affiliation[b]{Theoretical Physics Department, Fermilab, Batavia, Illinois 60510, USA}

\author[c]{Yonatan Kahn}

\affiliation[c]{Center for Theoretical Physics, Massachusetts Institute of Technology,\\Cambridge, MA 02139, USA}

\author[d]{and Matthew McCullough}

\affiliation[d]{Theory Division, CERN, 1211 Geneva 23, Switzerland}

\emailAdd{adama@mit.edu}
\emailAdd{pjfox@fnal.gov}
\emailAdd{ykahn@mit.edu}
\emailAdd{matthew.mccullough@cern.ch}

\date{\today}

\keywords{}

\arxivnumber{}

\abstract{Results from direct detection experiments are typically interpreted by employing an assumption about the dark matter velocity distribution, with results presented in the $m_\chi-\sigma_n$ plane.  Recently methods which are independent of the DM halo velocity distribution have been developed which present results in the $\vmin-\tilde{g}$ plane, but these in turn require an assumption on the dark matter mass.  Here we present an extension of these halo-independent methods for dark matter direct detection which does not require a fiducial choice of the dark matter mass.  With a change of variables from $\vmin$ to nuclear recoil momentum ($p_R$), the full halo-independent content of an experimental result for \emph{any} dark matter mass can be condensed into a single plot as a function of a new halo integral variable, which we call $\tilde{h}(p_R)$.  The entire family of conventional halo-independent $\tilde{g}(\vmin)$ plots for all DM masses are directly found from the single $\tilde{h}(p_R)$ plot through a simple rescaling of axes.  By considering results in $\tilde{h}(p_R)$ space, one can determine if two experiments are inconsistent for all masses and all physically possible halos, or for what range of dark matter masses the results are inconsistent for all halos, without the necessity of multiple $\tilde{g}(\vmin)$ plots for different DM masses.  We conduct a sample analysis comparing the CDMS II Si events to the null results from LUX, XENON10, and SuperCDMS using our method and discuss how the mass-independent limits can be strengthened by imposing the physically reasonable requirement of a finite halo escape velocity.}

\maketitle

\section{Introduction}
\label{sec:introduction}

The strong astrophysical evidence for dark matter (DM) has prompted an extensive program of experiments to detect DM terrestrially \cite{Bernabei:2010mq,Aprile2012,Behnke2012,Aguilar-Arevalo:2013uua,Agnes:2014bvk,Akerib2014,Agnese2014,Aalseth:2014eft,Angloher:2014myn}. Such direct-detection experiments exploit the possibility that DM can scatter off nuclei in a large detector volume \cite{Goodman:1984dc}, with a DM scattering event identified through the ensuing detection of the nuclear recoil energy $E_R$. However, the event rate as a function of $E_R$ is strongly dependent on the incident DM velocity distribution. In particular, the scattering rate is proportional to the integral over the entire velocity distribution of dark matter in the Galactic halo, which has never been measured directly. 

For simplicity it is often assumed that the speed distribution is Maxwell-Boltzmann in nature.  However, even in this simple case varying the Sun's circular velocity and the Galactic escape speed can significantly alter the interpretation of direct detection results \cite{Fairbairn:2008gz,MarchRussell:2008dy,McCabe:2010zh}. Similarly, marginalising over the parameters associated with the speed distribution motivated by dark matter numerical simulations, which show departures from pure Maxwell-Boltzmann, leads to uncertainty in interpretation of results \cite{Lisanti:2010qx,Mao:2013nda}.  Given the coarse resolution of numerical simulations and the uncertainties associated with the extracted speed distributions, one may instead parameterise the speed distribution in some general way \cite{Pato:2010zk,Peter:2011eu,Kavanagh:2012nr,Peter:2013aha}.  With sufficient data, and a judicious choice of basis for the parameterisation, one can hope to determine both the astrophysical and particle physics properties of dark matter simultaneously \cite{Kavanagh:2012nr,Kavanagh:2013wba}.
 
However, motivated by our present lack of understanding of the local dark matter velocity distribution, techniques have been developed which interpret direct detection results independent of assumptions about the astrophysics.  Given at least two experimental measurements of dark matter scattering, on different elements, it is possible to determine the dark matter mass without making assumptions about the form of the velocity distribution \cite{Drees:2008bv}.  Even without two positive results it is possible to compare experiments in an astrophysics independent fashion by constraining the halo integral itself, rather than the DM-nucleus scattering cross section \cite{Fox:2010bz}.  For a given nuclear recoil energy, only DM traveling at sufficiently high velocities can provoke a nuclear recoil at that energy. Thus, such halo-independent constraints are a function of the minimum DM velocity $v_{min}(E_R)$ required to provoke a recoil energy $E_R$, which for elastic scattering is
\be
v_{min} (E_R) = \sqrt{\frac{m_N E_R}{2 \mu_{N\chi}^2}} ~~.
\label{eq:vmin}
\ee
While we will limit our discussion to the case of elastically scattering dark matter in a single target experiment, these astrophysics-independent techniques have been extended to include multiple targets \cite{Frandsen:2011gi}, inelastic dark matter \cite{Bozorgnia:2013hsa}, momentum-dependent scattering \cite{Cherry:2014wia}, other more general forms of dark matter scattering \cite{DelNobile:2013cva}, searches for the modulating signal \cite{HerreroGarcia:2012fu}, and have been applied to compare data from many experiments \cite{McCabe:2011sr,Fox:2011px,Kelso:2011gd,Frandsen:2013cna,Fox:2013pia,Scopel:2014kba,DelNobile:2013cta,DelNobile:2013gba,Gelmini:2014psa}.  The techniques have also been extended to enable a halo-independent combined-likelihood function to be constructed \cite{Feldstein:2014ufa}, allowing for the  tension between DM hints and null results to be quantified in a systematic way, see also \cite{Bozorgnia:2014gsa}.

As can be seen from \Eq{eq:vmin}, $v_{min}$ depends explicitly on the DM mass $m_\chi$ through the DM-nucleus reduced mass $\mu_{N\chi}$. Thus, to analyze a direct detection experiment using existing halo-independent methods, one must make a choice for $m_\chi$. This is undesirable since it leads to a proliferation of plots, one for each choice of mass.  Furthermore, the goal of halo-independent methods is to ``factor out'' uncertainties about dark matter properties. Indeed, the DM mass is probably less well-constrained than the velocity distribution: while DM velocity in the Galactic halo is bounded by the Galactic escape velocity, there exist well-motivated DM candidates with masses spanning some twenty orders of magnitude \cite{Feng:2010gw}. Even in the range of masses $1 \ \GeV - 1 \ \TeV$ typically probed by direct detection experiments, different choices of DM mass can alter constraints derived from halo-independent analyses, and hints of positive signals can be either excluded or allowed in different regions of $v_{min}$ depending on the DM mass \cite{DelNobile:2013gba}.

In this paper we show that, for DM which undergoes non-relativistic elastic scattering with nuclei, many ambiguities related to the unknown DM mass can be resolved by a simple change of variables. By presenting limits and preferred values for the halo integral as a function of nuclear recoil \emph{momentum} $p_R$, instead of DM minimum velocity $\vmin$ or nuclear recoil energy $E_R$, one can condense the halo-independent results from an experiment in a way which is valid for \emph{any} value of the DM mass. This is a direct consequence of a kinematic relation for non-relativistic elastic scattering, derived from \Eq{eq:vmin},
\be
p_R = 2 \mu_{N\chi} v_{min} (E_R) = \sqrt{2 m_N E_R} ~~,
\label{eq:pnuc1}
\ee
which implies that $p_R$ is related to $v_{min}$ by a constant rescaling, but is manifestly independent of $m_\chi$.\footnote{This is the essential reason for working in $p_R$-space. A similar statement cannot be made about an $E_R$-space plot where, for nonrelativistic scattering, the rescaling of the $x$-axes into $v_{min}$-space is nonlinear in $E_R$ and explicitly depends on the DM mass. For relativistic scattering, though, one can make a mass-independent plot in $E_R$-space \cite{Cherry:2015oca}.} To put it another way, relating the observable $E_R$ to $v_{min}$ requires a choice for the DM mass, whereas relating $E_R$ to $p_R$ does not.\footnote{In \cite{Kavanagh:2012nr} it was also previously emphasized that the recoil momentum is an attractive variable for analyzing direct detection data, especially as it is independent of the DM mass.}  Thus, for a given experiment, the entire 1-parameter family of $\vmin$ plots for all DM masses can be condensed into a single $p_R$ plot, where the scale on the $p_R$-axis does not change as $m_\chi$ is varied.\footnote{This conclusion holds only for elastic scattering. The extent to which one can derive constraints without mass assumptions for more general kinematics is explored in a forthcoming paper \cite{FKMToAppear}.}  Furthermore, $p_R$ enjoys the same advantage of $\vmin$ in that multiple experiments with different nuclear targets, which may have different ranges of sensitivity in $E_R$, can in principle have overlapping sensitivity in $p_R$. 

While momentum-space plots are useful for a single experiment, the dependence of $p_R$ on the target mass $m_N$ means that care must be taken in comparing two different experiments on the same $p_R$ plot. Naively one would expect that the comparison must be carried out on a mass-by-mass basis, necessitating again that a large number of plots be made and compared.  However, we show that this issue can be avoided entirely by considering the \emph{ratios} of exclusion curves or envelopes of preferred values in $p_R$-space. By plotting these ratios as a function of a rescaled halo integral $\tilde{h}$, which does not depend on the target mass, a single plot contains sufficient information to determine if a dark matter signal is excluded for all dark matter halos \emph{and} for all dark matter masses, or whether there is a range of masses for which there is agreement. The rescaled halo integral $\tilde{h}$ has a conceptually straightforward interpretation as the differential scattering rate normalized to a single nucleon. In the most general case, it is simple to read off from this plot the range of DM masses for which two experiments are inconsistent for any DM halo.

This paper is organized as follows. In \Sec{sec:haloind}, we review the general setup of halo-independent methods, with particular emphasis on the role of $\vmin$. In \Sec{sec:unique}, we describe the change of variables from $\vmin$-space to $p_R$-space and show how this can be used to present experimental data in a halo-independent manner which is also independent of the DM mass.  The key results of this paper are in \Sec{sec:prefercomp}, where we show how $p_R$-space can be used to compare null results from one experiment with positive results from another experiment to derive regions of inconsistency which are valid for all DM masses above or below (depending on the ratio of target masses) a critical value. Finally, in \Sec{sec:example}, we perform a sample analysis on CDMS II Si, LUX, XENON10, and SuperCDMS data to illustrate the utility of $p_R$-space. For the convenience of the reader we summarize the minimal analysis recipe in \Sec{sec:recipe}. We conclude in \Sec{sec:conclusion}. In \App{app:limits} we comment on the use of mass-independent methods for comparing the relative strengths of different null results.

\section{Review of Halo-Independent Methods}
\label{sec:haloind}
Here we briefly review the standard techniques for halo-independent analyses of direct detection experiments. For spin-independent DM-nuclear scattering, the differential event rate at a direct detection experiment is given by
\be
\frac{dR}{d E_R} =  \frac{N_A \rho_\chi \sigma_n m_n}{2 m_\chi \mu_{n\chi}^2} C_T^2 (A,Z) \int d E'_R G (E_R,E'_R) \epsilon(E'_R) F^2 (E'_R) g(v_{min} (E'_R)) ~~,
\label{eq:ratea}
\ee
where $m_\chi$ is the DM mass, $m_n$ the nucleon mass, $\mu_{n\chi}$ the nucleon-DM reduced mass, $\sigma_n$ the DM-nucleon scattering cross-section, $\rho_\chi$ the local DM density, $N_A$ is Avogadro's number, $F (E_R)$ is the nuclear form factor which accounts for loss of coherence as the DM resolves sub-nuclear distance scales, $C_T(A,Z)= (f_p/f_n Z + (A-Z))$ is the usual coherent DM-nucleus coupling factor, $\epsilon (E_R)$ is the detector efficiency, and $G(E_R,E'_R)$ is the detector resolution function. 

The integral over the velocity distribution (which we will often refer to as the ``halo integral'') is
\be
\label{eq:gvmin}
g(v_{min}) = \int^\infty_{v_{min}} \frac{f(\mb{v}+\mb{v}_E)}{v} d^3 v ~~,
\ee
where $f(\mb{v})$ is the unknown DM velocity distribution and $\mb{v}_E$ is the Earth's velocity, both in the Galactic frame. We shall ignore the small time dependence of the Earth's velocity in the Galactic frame.  The lower limit of the halo integral, $v_{min}$, is the recoil energy-dependent minimum DM velocity required to produce a nuclear recoil $E_R$, and depends on the kinematics of the interaction. For elastic scattering $v_{min}(E_R)$ is given by \Eq{eq:vmin}.

As first pointed out in \cite{Fox:2010bu,Fox:2010bz}, since the DM velocity distribution $f(\mb{v})$ is positive semi-definite, $g(v_{min})$ must be a monotonically decreasing function of $v_{min}$. This leads to powerful constraints on the shape of the halo integral derived from either null or positive experimental results, which we will review shortly. First, though, we summarize the essential differences between the usual $m_\chi - \sigma_n$ analyses and halo-independent analyses.
\begin{itemize}
\item $\mb{m-\sigma}$ \textbf{plots}. Given a choice of DM velocity distribution $f(\mb{v})$ and DM local density $\rho_\chi$, event rates can be calculated as a function of $m_\chi$ and $\sigma_n$ from \Eq{eq:ratea}. Null results give exclusion contours, while positive results generally give closed preferred regions in $m_\chi - \sigma_n$ space. For each choice of $f(\mb{v})$, a different plot must be made.
\item \textbf{Halo-independent plots}. By rescaling the halo integral as
\be
\tilde{g}(\vmin) = \frac{\rho_\chi \sigma_n}{m_\chi} g(\vmin)~~,
\ee
to absorb all the detector independent terms in \Eq{eq:ratea}, event rates can be calculated from \Eq{eq:ratea} as a function of $\tilde{g}$, making the most conservative choice for $f(\mb{v})$. A choice for $m_\chi$ is still necessary to relate $v_{min}$ to $E_R$ through \Eq{eq:vmin}. Null results once again give exclusion contours, but in $\vmin - \tilde{g}(\vmin)$ space. Positive results give preferred shapes for the halo integral $\tilde{g}(\vmin)$, which is constrained to be a monotonically decreasing function; thus positive results generally give open \emph{envelopes} of such curves in $\vmin - \tilde{g}(\vmin)$ space, rather than closed regions. For each choice of $m_\chi$, a different plot must be made.
\end{itemize}
The choice of $\vmin$ as the $x$-axis variable for halo-independent plots is one of convenience, as it is the natural argument for the halo integral $\tilde{g}$, but since $\vmin(E_R)$ is monotonic for elastic scattering, one may just as easily present results in $E_R - \tilde{g}(E_R)$ space, since $\tilde{g}(E_R)$ is also monotonically decreasing. The advantage of $\vmin$ is that experiments with non-overlapping sensitivity in $E_R$ may have overlapping sensitivity in $\vmin$, due to the dependence of $\vmin$ on the target mass $m_N$. However, we will see in \Sec{sec:unique} that the choice of $p_R$ for the $x$-axis variable shares this same advantage, while at the same time obviating a choice of $m_\chi$.

To derive halo-independent exclusion curves, note that for a given point $v_0$ on the $\vmin$-axis, the halo integral which gives the least number of events consistent with monotonicity is
\be
\tilde{g}(\vmin) = \tilde{g}_0 \Theta(v_0 - \vmin)~~.
\ee
Setting an upper bound on the number of events using this choice of halo integral, one derives the weakest (\ie\ most conservative) possible bounds on $\tilde{g}(v_0)$. Thus, by sweeping over $v_0$ in the $\vmin$ range of an experiment, one can build up an exclusion contour which at every $x$-coordinate $v_0$ sets the most conservative bounds on the halo integral: the true halo integral must lie entirely below this exclusion contour for every $\vmin$. Such an exclusion curve is extremely robust, as any positive signal so excluded can be said to be excluded for \emph{all} DM halos. 

Determining best-fit values and envelopes for the halo integral for a positive result can be done in many ways. As a representative example, one may interpret \Eq{eq:ratea} as a component of an unbinned extended likelihood function
\be
\mathcal{L} = \frac{e^{-N_{E}}}{N_{O}!} \prod^{N_{O}}_{i=1} \frac{dR}{d E_R} \bigg|_{E_R=E_i}~~,
\ee
where $N_O$ total events are seen at recoil energies $E_i$.\footnote{For large $N_O$ the events may be binned.  In this case it is very simple to extract the preferred region of $\tilde{g}(\vmin)$ from binned likelihood or $\chi^2$ methods.} The monotonicity of $\tilde{g}$ can then be used to reduce the possible shapes of $\tilde{g}(\vmin)$ to a finite-dimensional subset, allowing a simple numerical maximization of the likelihood function. For details, we refer the reader to \cite{Fox:2014kua} where these statements are proven and expanded upon, as well as \cite{Feldstein:2014gza} which describes an alternate maximization technique. Including finite energy resolution effects, the preferred region at a given confidence level is an envelope in $\vmin - \tilde{g}(\vmin)$ space. 
Thus, for any point inside the envelope there is a curve which lies entirely within the envelope and which passes through that point which is consistent with the data at the required confidence.  In addition, there are no curves within the required confidence containing points which lie outside the envelope.
Consequently, an exclusion limit which crosses the lower boundary of this envelope is able to rule out \emph{all} DM halos preferred by the data, and thus excludes the putative signal in a halo-independent fashion.

When carrying out a halo-independent analysis, as mentioned above, a particular DM mass must be assumed.  The constraints on, or predictions for, $g(v)$ for another DM mass can be determined by simple rescalings \cite{Fox:2014kua}.  For a DM mass, $m_\chi$, a point $(v_{min},\tilde{g})$ is mapped to a new point $(v_{min}^\prime ,\tilde{g}^\prime)$ for DM mass $m_{\chi^\prime}$, by
\be
(v_{min}\,,\tilde{g}\,;m_\chi)\rightarrow \left(\frac{\mu_{N\chi}}{\mu_{N\chi^\prime}}v_{min}\,,\frac{\mu_{n\chi^\prime}^2}{\mu_{n\chi}^2}\tilde{g}\,;m_{\chi^\prime}\right)~.
\label{eq:gofvmapping}
\ee
Since the mapping depends on the mass of the nuclear target, this will shift different detectors by differing amounts.  In the following section we will present a change of variables which allows the presentation of results in a fashion independent of the DM mass.  Essentially this change of variables `factors out' the re-scaling freedom of \Eq{eq:gofvmapping} to leave a single plot which is the same for all $m_\chi$.

Of course, one could alternatively present results in the usual $\vmin - \tilde{g}(\vmin)$ space plot for a single choice of $m_\chi$ and then all other $\vmin - \tilde{g}(\vmin)$ space results for different masses could be immediately found from the rescaling of \Eq{eq:gofvmapping}.\footnote{This was described in more detail in \cite{Fox:2014kua}.}  However, there would be ambiguity surrounding which fiducial DM mass to take when presenting results in publications.  Furthermore, as we will see, the $p_R - \tilde{h}(p_R)$ space option advocated here allows for the potential inconsistency of two different experimental results for any halo and any DM mass to be determined directly by eye from the $p_R - \tilde{h}(p_R)$ space plot, whereas this is not possible for a $\vmin - \tilde{g}(\vmin)$ space plot with a specific choice of DM mass.

\section{Halo-Independent Direct Detection Without Mass Assumptions}
\label{sec:unique}
In \Sec{sec:method} we will describe the change of variables for calculating momentum-space constraints for a given experiment and then discuss how the standard halo-independent plot for any choice of DM mass may be easily constructed from this single plot.  The momentum-space plot is unique as only one such plot exists for any set of direct detection data, and is universal in the sense that it applies for all dark matter masses, and does not need to be reconstructed on a mass-by-mass basis, as is the case for the usual $v_{min}$-space plots. Following this, in \Sec{sec:prefercomp} we demonstrate the utility of the momentum-space plot by showing how the momentum-space constraints may in some cases be used to draw strong and general conclusions on the comparison between different direct detection experiments which hold for all DM halos and for all DM masses, making these conclusions halo-independent and true for all masses. Alternatively, the same information may, in some cases, imply a range of DM masses for which experiments are inconsistent, independent of DM halo. We show in \Sec{sec:escape} how these conclusions may be strengthened by restricting to halos with a finite escape velocity.

\subsection{Direct Detection in Momentum Space}
\label{sec:method}

\Eq{eq:ratea}, which describes the differential DM event rate, contains a number of unknown factors which depend on the DM mass, cross section, and local density.  To compare observations with specific models, one must make choices for these factors, but to simply compare results at different detectors this is unnecessary.  Thus, we may absorb all detector-independent factors into a single unknown, which we write as
\be
\tilde{h}(p_R(E_R))= \frac{N_A \rho_\chi \sigma_n m_n}{2 m_\chi \mu_{n\chi}^2} g(v_{min} (E_R)) 
= \frac{N_A m_n}{2 \mu_{n\chi}^2} \tilde{g}(\vmin) ~~.
\label{eq:Pdef}
\ee
We emphasize that the rescaling factor between $\tilde{h}$ and $\tilde{g}$ is \emph{independent} of the detector nucleus, since it only depends on the \emph{nucleon} mass $m_n$. As our notation makes clear, we are thinking of $\tilde{h}$ as a function of recoil momentum,
\be
p_R = \sqrt{2 m_N E_R}= 2\mu_{N\chi} \vmin~~,
\label{eq:pRdef}
\ee
which for a given recoil energy $E_R$ is \emph{independent} of DM mass, rather than $v_{min}$ which is not.  Both $\tilde{h}$ and $\tilde{g}$ are implicitly monotonically-decreasing functions of $E_R$, since $v_{min}$ is a monotonic function of $E_R$ and the halo integral $g$ is constrained by the arguments of \Sec{sec:haloind} to be a monotonic function of $v_{min}$. We can now rewrite \Eq{eq:ratea} as
\be
\frac{dR}{d E_R} =  C_T^2 (A,Z) \int d E'_R G (E_R,E'_R) \epsilon(E'_R) F^2 (E'_R) \tilde{h}(p_R(E'_R)) ~~.
\label{eq:rateanew}
\ee
Notice that in \Eq{eq:Pdef} we break with the convention of \cite{Fox:2010bz,Fox:2010bu}, and instead of the quantity $\tilde{g}$ we opt for $\tilde{h}$ which absorbs all dependence on the DM mass, and is also independent of the detector. The use of $\tilde{h}$ also has a conceptual advantage because it is a physical quantity: the differential scattering rate normalized to a single nucleon, before correcting for resolution and detection efficiency, analogous to the normalized WIMP-nucleon cross section used by experimental collaborations to report limits. This interpretation makes clear that halo-independent comparisons of experiments are just direct comparisons of recoil energy spectra, after a simple rescaling to account for the kinematic effects of different target masses.\footnote{We thank Felix Kahlhoefer for emphasizing this interpretation to us.}

Since there is a one-to-one relation between $\vmin$ and $p_R$, it is simple to repeat the discussion of \Sec{sec:haloind} in terms of $p_R$.  For instance, if DM scattering were to produce an event with nuclear recoil momentum $p_0$, then the halo integral which could lead to this event while producing the minimum number of scattering events at other recoil momenta is described by
\be
\label{eq:PTheta}
\tilde{h} (p_R) = \tilde{h}_0 \Theta(p_0-p_R)~~.
\ee
Here, $p_0$ is taken over the whole range of $p_R$ sensitivity of the experiment: if the experiment is sensitive to energies satisfying $E_{low}\le E \le E_{high}$, then $\sqrt{2m_N E_{low}} \le p_R \le \sqrt{2m_N E_{high}}$. Thus, one can set limits on $\tilde{h}_0$, and thus $\tilde{h}(p_0)$, based on the null results of an experiment by integrating over $E_R$ and \eg\ using a Poisson upper limit on the total number of events at a given confidence level, for each value of $p_0$. Just as easily, $\tilde{h}(p_R)$ can be chosen to be a sum of step functions as proposed in \cite{Fox:2014kua,Feldstein:2014gza} in order to find best-fit regions for the halo integral in the case of positive signals. In either case, by construction this analysis makes only the most conservative assumptions about the DM velocity distribution.  We emphasize that at no point in calculating the constraints on \Eq{eq:rateanew} is it necessary to make a choice for $m_\chi$.

It is convenient at this stage to make a few observations on the possible form of momentum-space exclusion curves which will be useful later.
\begin{itemize}
\item  As a detector is not sensitive to nuclear recoils at vanishing momentum, and most analyses have a low-momentum threshold, the exclusion curves will in general become weaker at low recoil momenta and asymptote to infinity at some fixed small recoil momentum.
\item  As the most conservative choice of halo (\ref{eq:PTheta}) predicts larger numbers of events for larger recoil momenta, the limits will in general become stronger towards higher recoil momenta. The result is that the exclusion curve in momentum-space will in general be a monotonically decreasing function.\footnote{Although $\tilde{h}$ should, on general grounds, be a monotonically decreasing function, the limits on $\tilde{h}$ extracted from real data may not be monotonically decreasing.  For example, if Yellin's methods \cite{Yellin:2002xd} are used, then limits can increase near the $v_{min}$ where the limit-setting method switches from one interval to another.} 
\item  Detectors in principle have sensitivity out to infinite recoil momenta, as a signal beyond the high momentum threshold of a detector implies a nonzero number of predicted events within the signal region. Thus, the momentum space exclusion contour will in principle extend to infinite recoil momenta, but due to the finite energy range of the signal region, the bound will approach a constant value in $\tilde{h}$.
\end{itemize}

The momentum-space representation of experimental results contains all of the information conveyed by the usual $v_{min}$-space plots, and these plots may actually be constructed directly from one another.  From \Eqs{eq:Pdef}{eq:pRdef}, the mapping is linear for any choice of DM mass:
\be
\left({p_R},\tilde{h} \right) \mapsto \left(\frac{1}{2 \mu_{N\chi}} p_R,  \frac{2 \mu_{n\chi}^2}{N_{A} m_n} \tilde{h} \right) = ({v_{min}},\tilde{g}) ~~.
\label{eq:map}
\ee
From the behaviour of $\tilde{g}(v)$ under changes of DM mass, \Eq{eq:gofvmapping}, we see that the single momentum-space plot contains all halo-independent information.  

\begin{figure}[t]
  \centering
 \includegraphics[height=0.35\textwidth]{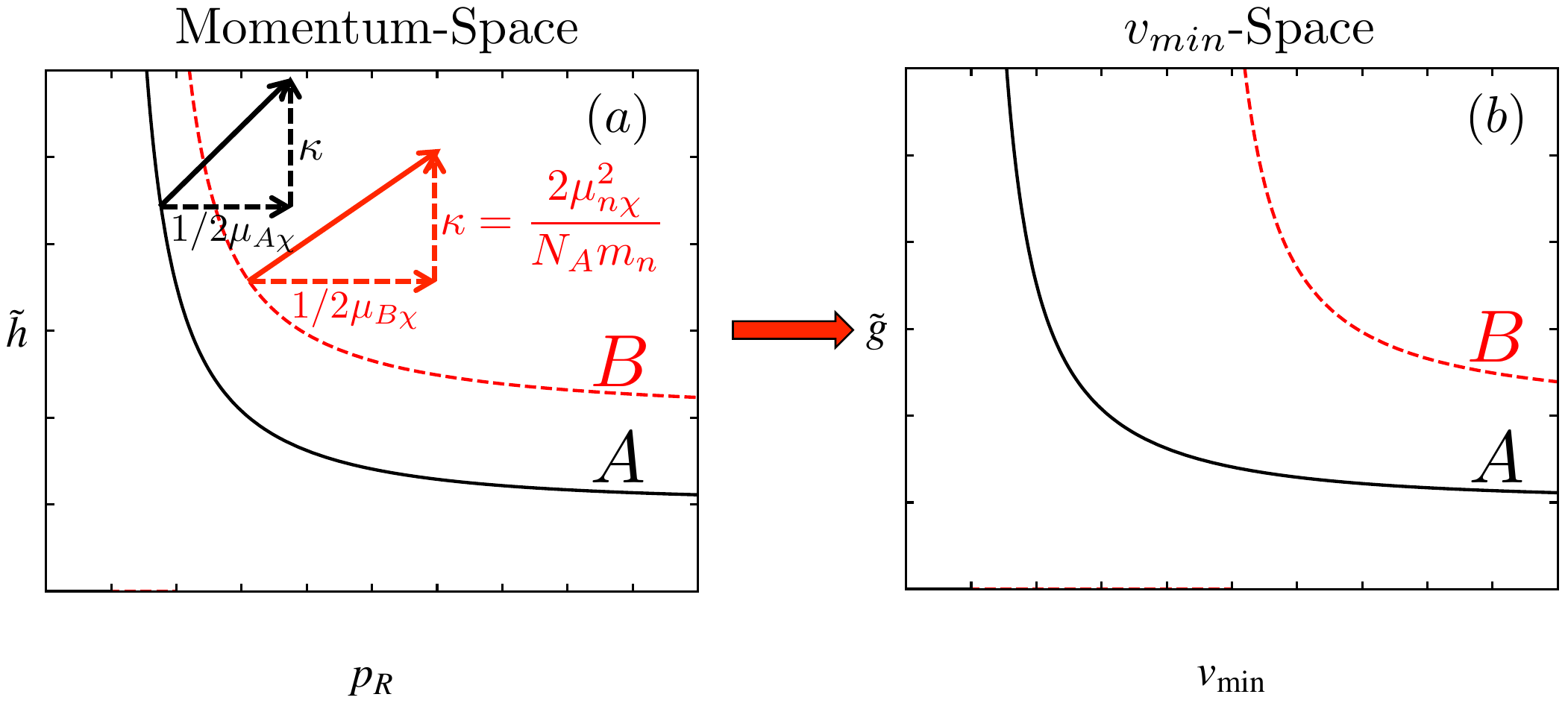} 
\caption{(a) A schematic of the momentum-space plot proposed in this work for two different detectors with target nuclei masses satisfying $M_A>M_B$.  (b)  A schematic of the standard $v_{min}$-space plot.  The mapping from the unique momentum-space plot to the $v_{min}$-space plot is also shown in (a) for a typical choice of DM mass making clear that the differences only arise in the horizontal direction, see \Eq{eq:map}.}
  \label{fig:mapping}
\end{figure}

We can also compare the results of two different experiments on the same $p_R$ plot. For any DM mass, the mapping of \Eq{eq:map} rescales both curves in the vertical direction by the \emph{same} factor, but rescales them in the horizontal direction by different factors depending on both the DM mass and the target nucleus mass. Hence, in mapping from recoil momentum space to $v_{min}$-space, exclusion curves for two different detectors are shifted relative to each other \emph{only} in the horizontal direction, as depicted in \Fig{fig:mapping}. We now show how this may be exploited to draw very general halo-independent conclusions on the relative behavior of different experiments which hold for all DM masses.

\subsection{Comparisons Between Null and Positive Results}
\label{sec:prefercomp}
\begin{figure}[t]
  \centering
 \includegraphics[height=0.41\textwidth]{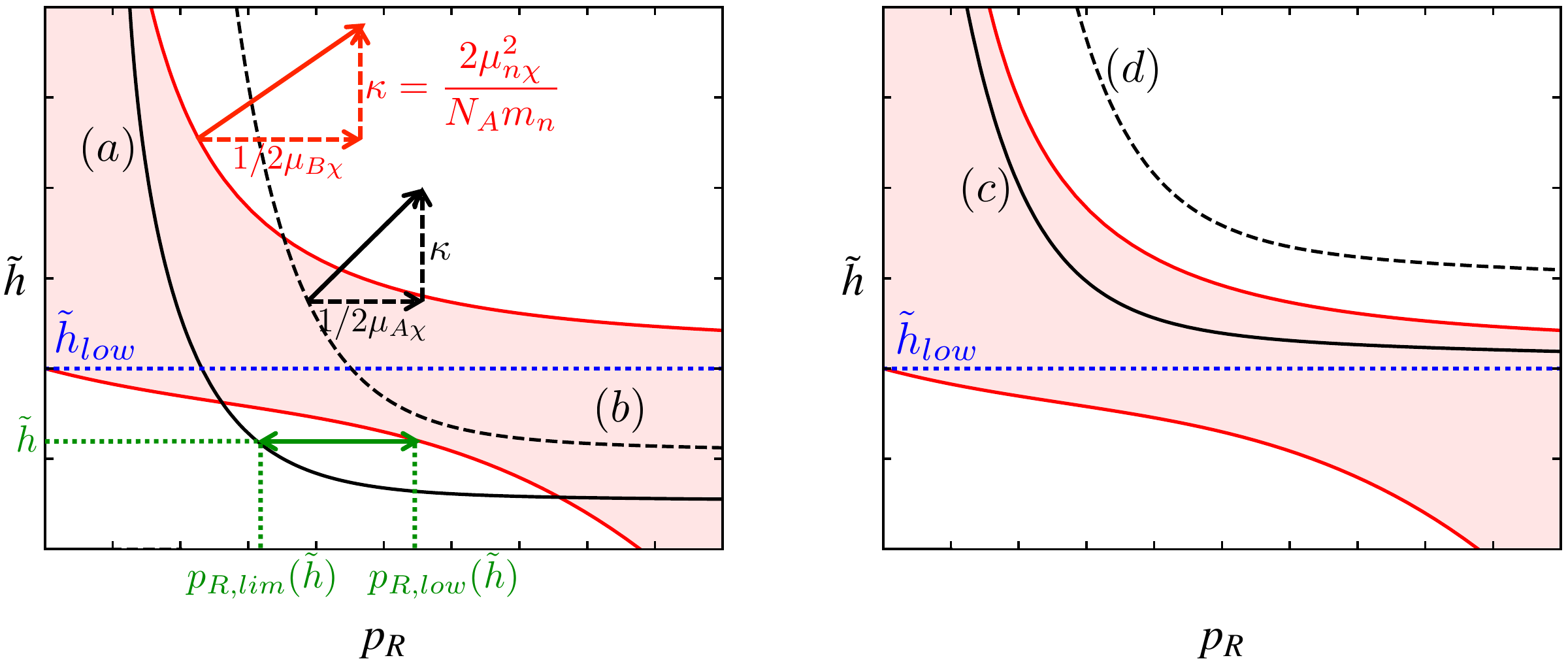} 
\caption{Illustrative possibilities of momentum-space plots for results from an experiment $B$ (shaded red) which observes a tentative DM signal and a null experiment $A$ (black lines) which is used to set constraints.  The target nuclei masses are here assumed to satisfy $M_A>M_B$. The rescaling to map from momentum-space to $v_{min}$-space is only shown in one case for clarity, but this mapping holds for all cases.  In (a) an experiment $A$ finds constraints which exclude a DM interpretation of all anomalous events observed in $B$ as the exclusion contour crosses the lower boundary of the preferred region. Any mapping to $v_{min}$-space will shift the region for $B$ further to the right relative to $A$, implying that experiment $A$ excludes a DM interpretation of $B$ for all DM masses and all DM halos. In (b) it may be that $A$ excludes a DM interpretation of $B$ for all halos for some finite range of DM masses. In (c) and (d), experiment $A$ can never exclude a DM interpretation of $B$ for all halos for any choice of DM mass, as shifting $B$ to the right never causes $A$ to cross the lower boundary of $B$. However, in (c) there is tension between the results of $A$ and $B$ independent of the DM mass, whereas in (d) there is only tension for a range of DM masses. See the text for a general method to resolve and quantify these differences.  In green we also show the construction of the variable $R(\tilde{h})$, defined in \Eq{eq:RPref} which allows for a quantitative assessment of all possibilities.}
  \label{fig:schematicprefer}
\end{figure}

A principal purpose of the momentum-space method will be to serve as a general tool to determine the consistency of potential positive results with bounds from detectors which observe no hint of DM scattering.  There is a great deal of value in this sort of comparison, as it may shed light on the nature of an emerging DM discovery and possibly point towards DM mass ranges which are preferred for such a signal. In general, positive results give preferred regions which explain the observed scattering events in a detector to within some given statistical confidence.  Technology for determining these best-fit regions has been developed in the context of halo-independent $v_{min}$-space plots and may be easily adapted for momentum-space plots.  As an example, using the method described in \cite{Fox:2014kua} it is possible to determine the preferred parameter space using an unbinned likelihood function, leading to preferred regions such as those depicted schematically in \Fig{fig:schematicprefer}. Unlike the preferred regions in $m_\chi - \sigma_n$ space, the preferred parameter regions in $\vmin$-space will not in general be closed due to the monotonicity constraint on the halo integral.

At this point it is worth discussing some general features of the preferred parameter regions in momentum space.  The upper boundary of a preferred region will be very similar to an exclusion contour.  Signal events are preferred, but not too many; this boundary acts as an exclusion contour which may extend to non-zero values for arbitrarily high recoil momenta. It also becomes weak below the threshold of the detector, approaching infinity as the recoil momentum tends to zero.  The lower boundary of a preferred region is different.  Above the high energy threshold of a detector there is no preference for signal events, thus the lower boundary will tend towards zero at high recoil momenta.  At recoil momenta below the low energy threshold there is also no preference for signal, however monotonicity combined with the preference for signal at higher momenta means that the lower boundary of the preferred region will approach a fixed value at low recoil momenta, which we denote $\tilde{h}_{low}$.  All of these features are depicted in \Fig{fig:schematicprefer}; it is worth emphasizing that these features would be expected for a momentum-space plot for any DM hint.

We consider two experiments: the null experiment $A$ which may be used to place constraints in momentum-space, and experiment $B$ which observes anomalous scattering events which may be interpreted as DM scattering. In the schematic of \Fig{fig:schematicprefer}, the black curves are from $A$ and the red shaded regions are from $B$. In some simple cases it is possible to determine the relative strengths of bounds from two different detectors directly from their $\tilde{h} (p_R)$ plots as described in the caption.

In more general cases, determining the outcome of mapping the momentum-space exclusions to $v_{min}$-space for a specific DM mass is less straightforward. To treat the general case we exploit the fact that mapping from momentum-space to $v_{min}$-space only shifts curves relative to each other in the horizontal direction.  Each momentum-space exclusion curve or lower boundary of a preferred region is given by a function $\tilde{h} (p_R)$; since this curve is monotonically decreasing, we may define the inverse mapping $p_R (\tilde{h})$ over the fixed range 
\be
\tilde{h} \in [\tilde{h} (p_R \to \infty),\infty]~~.
\label{eq:hmap}
\ee
In order to quantify the relationship between exclusion contours and preferred regions of parameter space it is useful to consider ratios of $p_R$ as a function of $\tilde{h}$.  Specifically, we define
\be
R(\tilde{h}) = \frac{p_{R,\text{lim}} (\tilde{h})}{p_{R,\text{low}} (\tilde{h})}  ~~,
\label{eq:RPref}
\ee
where $p_{R,\text{lim}}$ is an exclusion curve, and $p_{R,\text{low}}(\tilde{h})$ is the curve describing the lower boundary of a preferred region.\footnote{In some cases, such as (c) and (d) of \Fig{fig:schematicprefer}, $R$ is not well-defined as there is no overlap in the regions of $\tilde{h}$ explored by the exclusion contour and the lower boundary of the preferred region. In these cases it is clear by eye that there are no DM masses for which $A$ can exclude $B$ independent of the DM halo.} This construction is shown in green in \Fig{fig:schematicprefer}.  In the case of a positive signal, it is the lower boundary of the preferred region which determines the consistency between a DM interpretation of experiment $B$ relative to an exclusion from experiment $A$. We also define a related quantity
\be
R_\chi (\tilde{h}) = \frac{v_{min,\text{lim}} (\tilde{h})}{v_{min,\text{low}} (\tilde{h})}~~,
\ee
which is the ratio of the two curves in $\vmin$ space.\footnote{Since the mapping from $\tilde{h}$ to $\tilde{g}$ is the same for both $A$ and $B$, we prefer to think of $R_\chi$ as a function of $\tilde{h}$ rather than $\tilde{g}$ for simplicity of notation.}  As the notation makes clear, $R_\chi$ depends explicitly on $m_\chi$. If $R_\chi (\tilde{h}) < 1$, the exclusion contour from $A$ crosses below the lower boundary of the preferred region from $B$, and we can say that the signal from $B$ is inconsistent with the null results from $A$ for the particular mass $m_\chi$ which defines the mapping into $\vmin$-space.

\begin{figure}[t]
  \centering
  \includegraphics[width=0.95\textwidth]{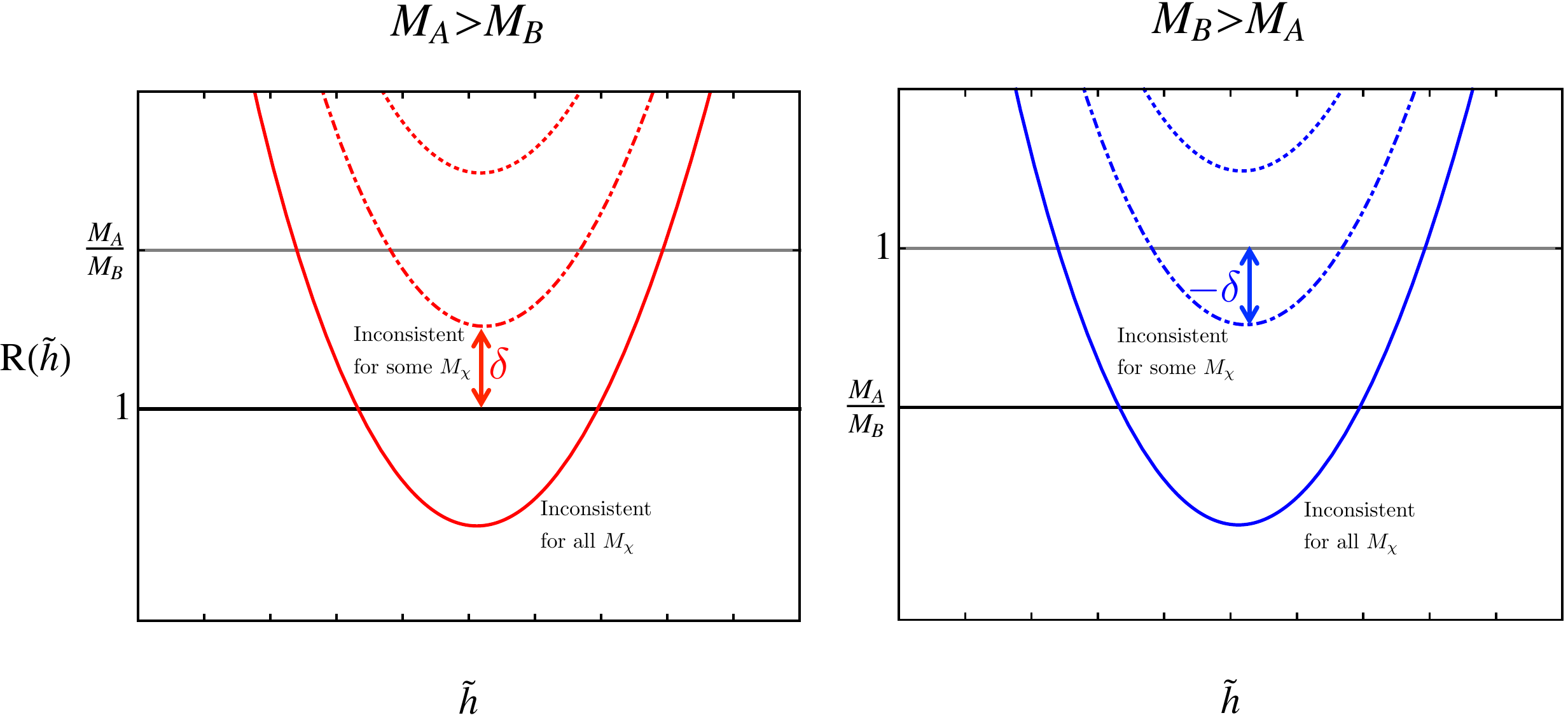} 
\caption{Some possibilities for the variable \Eq{eq:RPref}, for both possible mass orderings. When moving from $p_R$-space to $v_{min}$-space, the height of any line is rescaled by the DM mass-dependent parameter \Eq{eq:rat} which satisfies the appropriate inequality in \Eq{eq:ineq}. Case (a) in \Fig{fig:schematicprefer} corresponds to the solid red curve, and case (b) corresponds to the dashed or dot-dashed red curves. Cases (c) and (d) cannot be plotted because the lower boundary and the exclusion curve have no overlap in $\tilde{h}$, and $R(\tilde{h})$ is undefined.}
  \label{fig:ratioprefer}
\end{figure}

The relationship between $R$ and $R_\chi$ is a simple scaling,
\be
R_\chi = F_\chi R~~,
\ee
with
\be
F_\chi = \frac{\mu_{B\chi}}{\mu_{A\chi}}  =   \frac{M_B}{M_A} \frac{M_A+M_\chi}{M_B+M_\chi} ~~.
\label{eq:rat}
\ee
This scaling is inherited from the scaling of $p_R$ given in \Eq{eq:map}. Crucially, over all DM masses $F_\chi$ satisfies the inequality
\be
 \left\{ \begin{array}{lr}
\frac{M_B}{M_A} \le F_\chi \le 1, \ {\rm if\ } M_A > M_B \\
1 \le F_\chi \le \frac{M_B}{M_A}, \  {\rm if\ } M_B > M_A~~.
\end{array}
\right.
\label{eq:ineq}
\ee
This restricted range of $F_\chi$ is a key tool for extracting mass-independent statements concerning different experiments. Consider the case of $M_A>M_B$. If $R$ lies entirely above $M_A/M_B$, then the two results are consistent for all DM masses, since as the DM mass is varied, $R_\chi$ can never dip below 1 thanks to the inequality (\ref{eq:ineq}). If any of the $R$ curve lies below 1, then $A$ and $B$ are inconsistent for all DM masses since some point on $R_\chi$ will lie below 1 for all $m_\chi$. Similar reasoning holds for the case $M_A < M_B$, but with the roles of the critical lines at 1 and $M_A/M_B$ reversed; see \Fig{fig:ratioprefer}.

Suppose the global minimum of $R(\tilde{h})$ is located at $\tilde{h} = \tilde{h}_0$. In the case where $R(\tilde{h}_0)$ lies between 1 and $M_A/M_B$ (for either mass ordering), define
\be
\delta = R(\tilde{h}_0) - 1~~.
\label{eq:deltadef}
\ee
Note that $\delta < 0$ for the case $M_B > M_A$. Depending on $\delta$, there will be a range of DM masses for which $A$ excludes $B$. The lower or upper endpoint of that range (depending on the mass ordering) will be the value of $M_\chi$ where $R_\chi(\tilde{h}_0) = 1$, given by 
\be
M_{AB}(\delta)=  \frac{M_A M_B \delta}{M_A-(1+\delta) M_B}~~.
\label{eq:MAB}
\ee
For the two cases the range of DM masses for which $A$ excludes $B$ is
\be
\label{eq:MchiLimited}
\left\{ \begin{array}{lc}
M_\chi >  M_{AB}(\delta), \ {\rm if\ } M_A > M_B \\
M_\chi <  M_{AB}(\delta), \  {\rm if\ } M_B > M_A~~.
\end{array}
\right.
\ee
A schematic of these various scenarios is shown in \Fig{fig:ratioprefer}. 
 
We emphasize that all of the results discussed above hold for all DM halos because the constraints are constructed for the most conservative halo possible. Furthermore, carrying out the mapping into $\vmin$-space to obtain $R_\chi$ is actually unnecessary, since the $R(\tilde{h})$ plot contains all the necessary information. Thus in practice, the entire analysis can be carried out without a choice of $m_\chi$ by working exclusively with $R(\tilde{h})$.\footnote{The mapping into $\vmin$-space in this section was performed to make explicit the connection between previous methods and this method.}

We focus here on pairwise comparisons of one hint (defined by a best fit region at confidence level $CL_{hint}$) with one exclusion (defined by a bound at confidence level $CL_{bound}$) at a time. It would be straightforward to combine all hints in a joint likelihood function and compare this region with each constraint individually.  We do not attempt to quantify the level of tension, if there is any, beyond the statement that the two results are inconsistent at their respective confidence levels.  Such techniques do exist \cite{Bozorgnia:2014gsa}, but we have not yet tried to extend them to $\tilde{h}$-space.  We have assumed the bounds are derived using techniques that do not require the form of the background be known, such as the techniques of Yellin \cite{Yellin:2002xd}.  However, if the background is well-modeled one can go further and combine both signals and bounds into one global likelihood function and compare the relative likelihood of various hypotheses to fit all the data \cite{Feldstein:2014ufa}.

\subsection{Additional Constraints from Escape Velocity}
\label{sec:escape}

Up to this point, our analysis has been completely halo-independent, in the sense that we have imposed no constraints on the DM velocity distribution, no matter how physically reasonable. Astrophysical halos, however, have escape velocities, and it is worth considering the effect of imposing a cutoff on the velocity distribution at $v_{esc}$ on the behavior of momentum-space exclusion limits. In $v_{min}$-space, a cutoff at $v_{esc}$ means that a preferred region for a positive signal at $v_{min} > v_{esc}$ should no longer be interpreted as coming from DM scattering. Consider again the relation between $\vmin$ and $p_R$,
\be
v_{min} = \frac{p_R}{2 \mu_{N\chi}} ~~.
\label{eq:vminpR}
\ee
Thanks to the reduced mass in the denominator, the rescaling to $v_{min}$-space is a monotonically decreasing function of $m_\chi$, which means that for sufficiently small $m_\chi$, a given point on the $p_R$-axis will always be mapped to $v_{esc}$:
\be
m_{\chi, min} = \frac{m_N p_R}{2m_N v_{esc} - p_R} \implies p_R \mapsto v_{esc}~~.
\label{eq:mapstovesc}
\ee
For this $p_R$ value, only $m_\chi$ larger than $m_{\chi,min}$ can be consistent with a DM interpretation.

As a consequence, halo-independent exclusion limits which hold only for limited DM mass ranges may in fact hold over all DM masses consistent with physically reasonable halos. Suppose that $M_A > M_B$, and let $p_{B,max}$ be the largest $p_R$-coordinate of the lower boundary of envelope $B$. Setting $m_N = M_B$ in \Eq{eq:mapstovesc} and comparing with \Eq{eq:MchiLimited}, one can conclude that $A$ excludes $B$ for all DM halos with escape velocity less than or equal to
\be
v_{esc} = \frac{(M_A - M_B)p_{B,max}(1+\delta)}{2 M_A M_B \delta}~~.
\label{eq:vescCrit}
\ee
Put another way, the minimum DM mass required to bring $A$ and $B$ into agreement is already small enough that every halo in the preferred region implies $\vmin$ values greater than this $v_{esc}$. Note that such an analysis is not possible when $M_B > M_A$, because in that case $F_\chi \geq 1$ and the relevant inequalities are reversed. However, if $m_{\chi, min} > M_{AB}(\delta)$, then the constraints on DM masses from the escape velocity are stronger than the requirements for consistency between results from $R(\tilde{h})$. 

\section{A Sample Analysis: Comparing CDMS II Si with SuperCDMS, LUX, and XENON10}
\label{sec:example}
To demonstrate the methods of section \ref{sec:prefercomp} for comparing positive signals with exclusions, we consider the case of the SuperCDMS, LUX, and XENON10 limits and the preferred halo for the Si detectors of the CDMS II experiment.

The SuperCDMS collaboration has publicly released data from their analysis of 577~kg~d of exposure on their seven Ge detectors with the lowest energy thresholds \cite{Agnese2014,SuperCDMSData2014}. The blind analysis observed 11 events with $6.1^{+1.1}_{-0.8}$ events expected from background. Although the observed events are consistent with the background estimate, the SuperCDMS collaboration did not perform a background subtraction before calculating the limit, conservatively assuming that all observed events are potential DM scattering. Using the efficiency, ionization yield, and energies of the 11 events from the data release, we set 90\% C.L. limits using the ``Pmax" method \cite{Yellin:2002xd}.

The CDMS II Si analysis observed three events in 140.2~kg~d of exposure. We follow the analysis approach used in \cite{Fox:2014kua}, taking a gaussian detector resolution of 0.2~keV, the acceptance from \cite{Agnese2013}, and backgrounds from \cite{APStalk}. We furthermore assume that the distribution of $\Delta L$ is $\chi^2$ corresponding to five degrees of freedom: one for each step position and height in the halo, minus one for the monotonicity constraint. The $\chi^2$ assumption is a reasonable approximation to the sampling distribution of $\Delta L$ \cite{Fox:2014kua}, a more precise determination could be made by 
a Monte Carlo simulation, varying the number of events in each pseudoexperiment. A value of $\Delta L = 9.2$ then corresponds to 90\% C.L. We perform Markov Chain Monte Carlo (MCMC) sampling\footnote{A computer code developed to determine these envelopes in general is available from the authors upon request.} of the 6-dimensional parameter space, and use halos with $\Delta L < 9.2$ to calculate the envelope preferred by the CDMS II Si result.

For comparison, we also compute limits for LUX and XENON10 at a fixed dark matter mass as described in section \ref{sec:haloind}. Similar to \cite{Fox:2014kua}, we use events and ionization yield $\mathcal{Q}_y$ from the XENON10 S2-only analysis \cite{Angle2011}. We conservatively assume that the ionization yield drops to zero below 1.4~keVnr because the signal acceptance cannot be reliably estimated below this energy. The energy resolution is taken to be $\Delta E_R = E_R / \sqrt{E_R \mathcal{Q}_y(E_R)}$, while the acceptance is 95\% on an exposure of 15~kg~d. Yellin's ``Pmax'' method \cite{Yellin:2002xd} is used to set limits at 90\% C.L., using the conservative assumption that all observed events are potential DM scatterings. Following \cite{Fox:2014kua}, we deduce Poisson upper limits from the results of the LUX experiment \cite{Akerib2014}. The background distributions for the LUX experiment have not been made public, so we cannot compute limits based on the profile likelihood ratio test statistic as done by the LUX collaboration. Because no events were observed in LUX in the low-energy nuclear recoil band, we instead use the 90\% upper limit for an observation of 0 events for a Poisson process with no background, taking the efficiency from \cite{Akerib2014}. This simplification gives reasonable agreement with the limits reported by LUX in the low-mass region.\footnote{See \cite{DelNobile:2013gba,Fox:2013pia} for more details on reconstructing the published LUX results.}

\begin{figure}[]
\centering
{\includegraphics[width=3.0in]{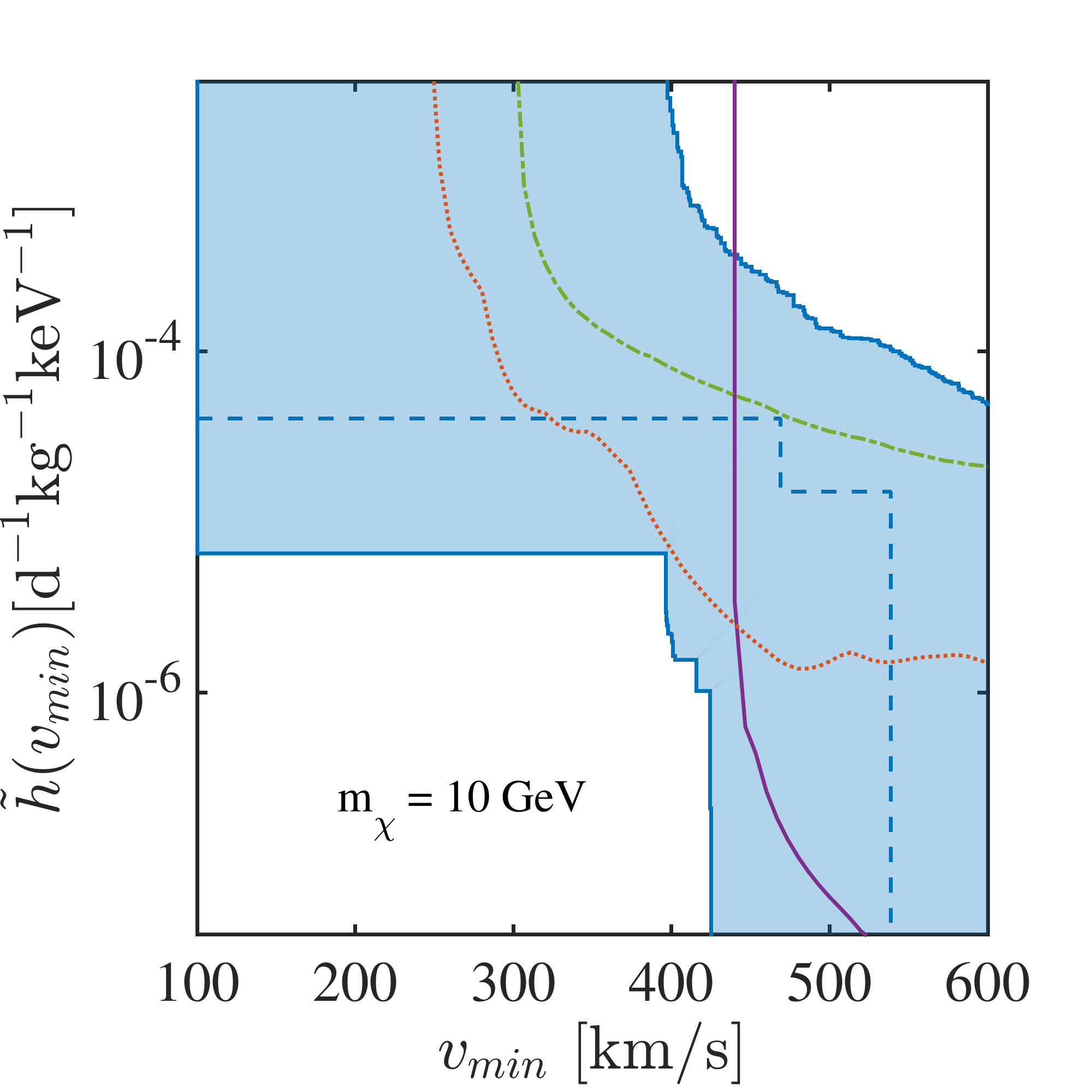}}
{\includegraphics[width=3.0in]{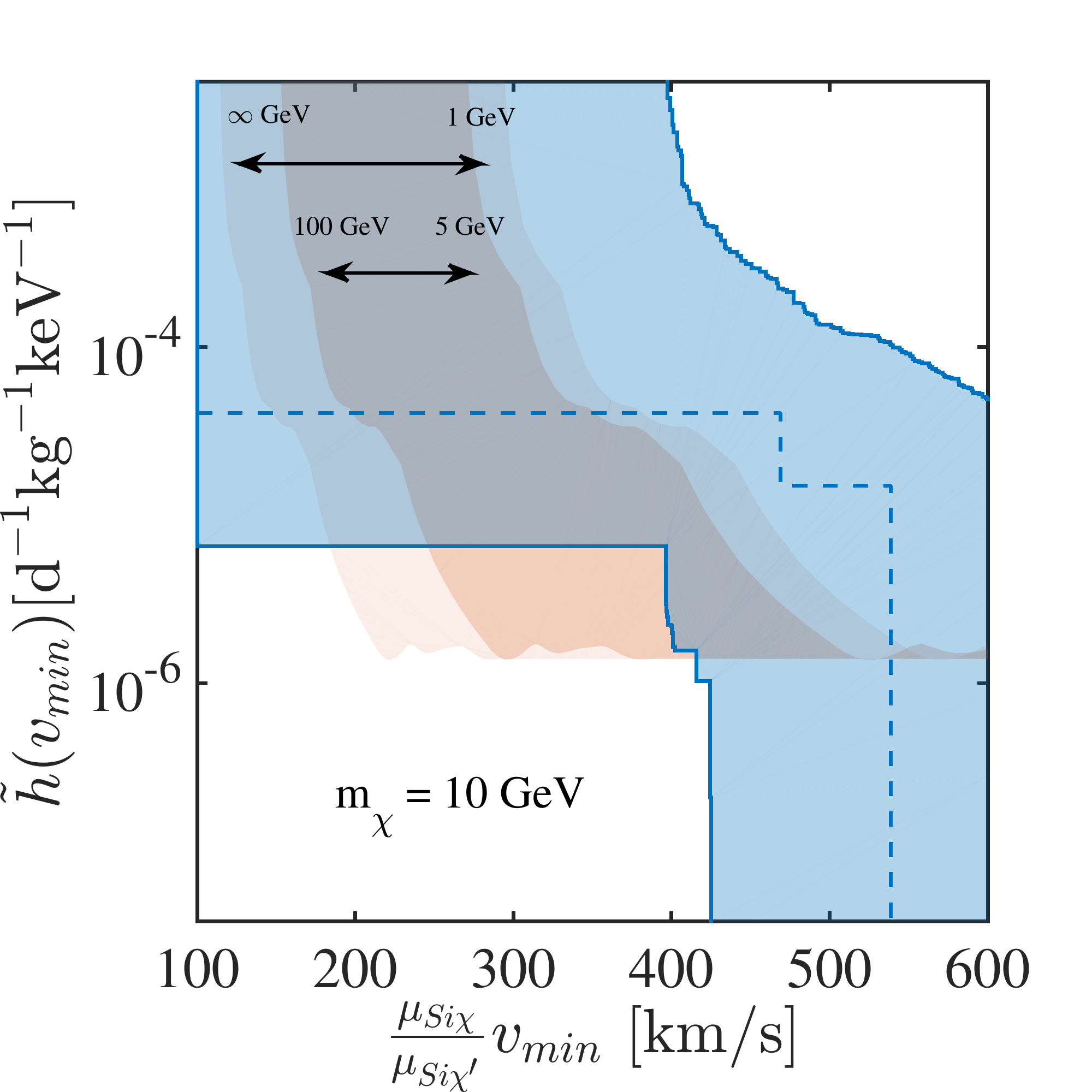}}
\caption{Exclusion limits from XENON10 (dot-dashed green), LUX (solid purple), SuperCDMS (dotted orange) overlaid on the preferred region (90\% C.L.) for CDMS II Si (shaded blue) and best fit halo (dashed blue) in $v_{min}$-space. {\bf Left}: $\vmin$-space plot for 10~GeV DM. {\bf Right}: range of SuperCDMS limits scaled with respect to 10~GeV DM using the factor appropriate for Si (shaded orange) (analogous to Fig. \ref{fig:vMinLimits}, right), overlaid with the preferred region for CDMS II Si (shaded blue).}
\label{fig:vMinContour}
\end{figure}

In the left panel of Fig. \ref{fig:vMinContour} we compare limits from XENON10, LUX, and SuperCDMS, for 10~GeV DM in the usual $v_{min}$-space plot. As pointed out in \cite{Fox:2014kua}, the limits in $\vmin$-space exhibit a simple scaling with the DM mass,
\be
v'_{min} (E_R) = \frac{\mu_{N\chi}}{\mu_{N\chi'}} v_{min} (E_R) ~~.
\label{eq:mapvmin}
\ee
This can be visualized in a single plot, such as the right panel of Fig. \ref{fig:vMinContour}, which shows the range of limits for SuperCDMS, when referred to 10~GeV DM using the scaling factor appropriate for Ge. Such a plot will contain, for every possible mass, the correct positioning of the CDMS-Si preferred region relative to SuperCDMS. For sufficiently large DM masses, the exclusion limit from SuperCDMS drops below the lower boundary of the CDMS II Si contour, indicating strong tension between the inferred dark matter halos of the two experiments. The mass at which the SuperCDMS exclusion drops below the lower limit of the CDMS II Si preferred region can be assessed using the methods of \Sec{sec:prefercomp}.

\begin{figure}[]
  \centering
\includegraphics[width=3.0in]{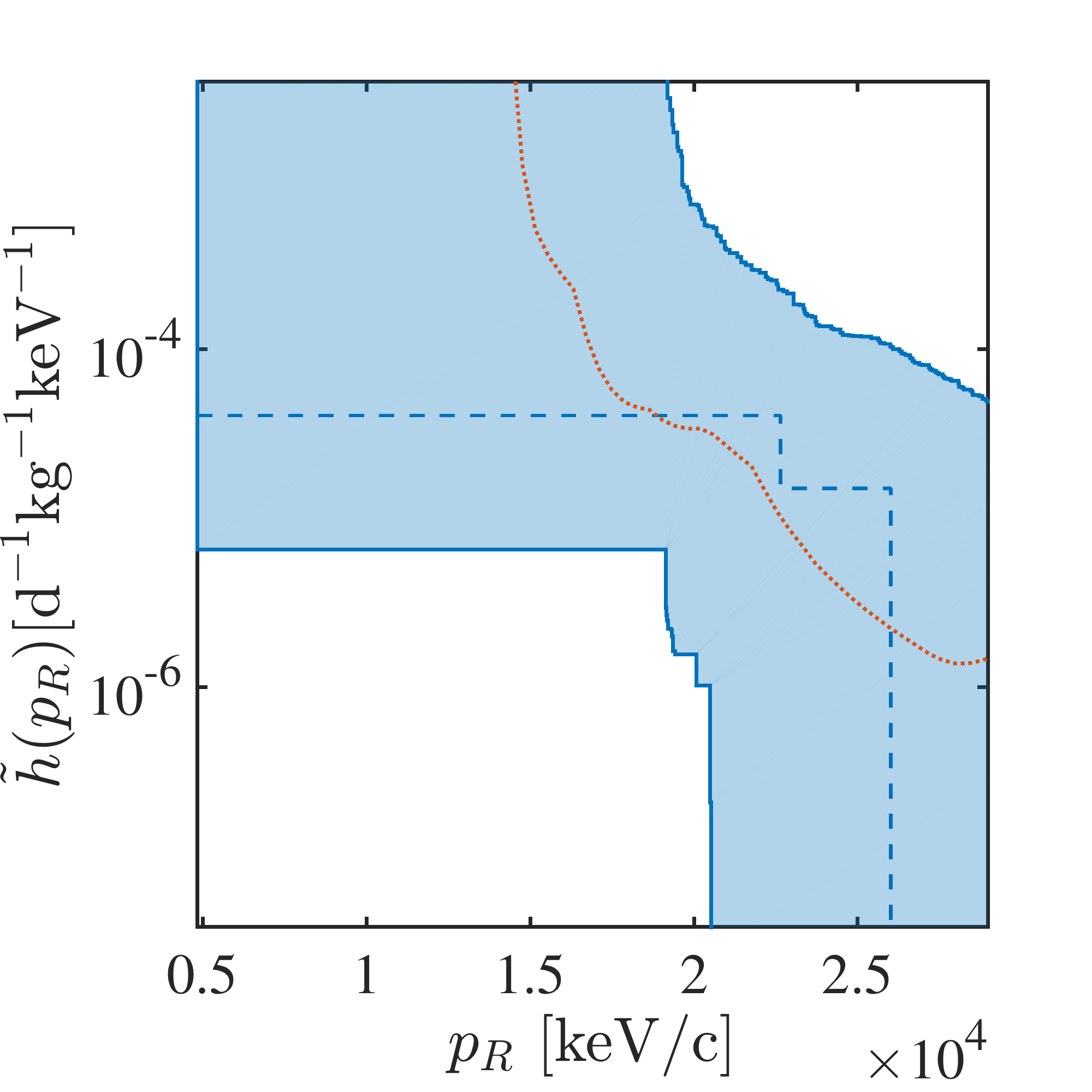}
\includegraphics[width=3.0in]{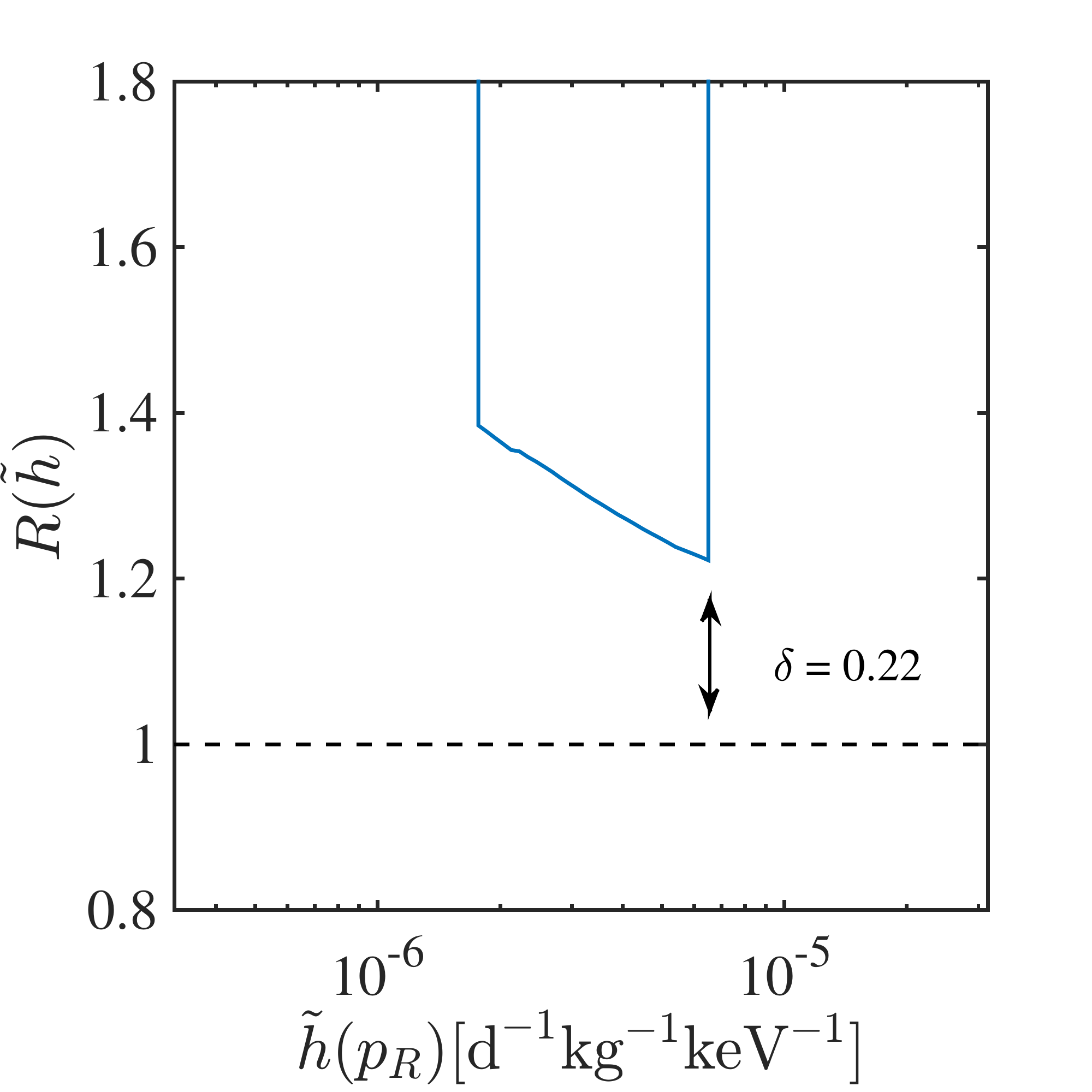}
\caption{Momentum-space plot comparisons between CDMS II Si and SuperCDMS. {\bf Left}: exclusion limits from SuperCDMS (dotted orange) in $p_R$-space overlaid with preferred region of CDMS II Si (shaded blue) and best fit halo (dashed blue). {\bf Right}: $R$-ratio of SuperCDMS exclusion with lower boundary of CDMS II Si preferred region. Deviations from monotonicity are smoothed so that the limits $\tilde{h}(p_R)$ can be inverted to form $R(\tilde{h})$, as in equation (\ref{eq:RPref}). The minimum value of $\delta = 0.22$, implies that SuperCDMS excludes CDMS II Si across the full range of $v_{min}$ only for $m_\chi \gtrsim 11$~GeV. For halos with escape velocity less than $v_{esc} \simeq 410 \ {\rm km}/{\rm s}$, SuperCDMS excludes CDMS II Si for all DM masses. The critical horizontal line at $M_{{\rm Ge}}/M_{{\rm Si}}$ is above the top of the plot.}
\label{fig:pRContour}
\end{figure}

The same information is more directly accessed in the corresponding momentum-space plots, shown in Fig. \ref{fig:pRContour}, which have the advantage of not requiring a choice of reference DM mass. Because the lower boundary of the CDMS II Si region and the SuperCDMS exclusion have some overlap along the $\tilde{h}$-axis, their comparison corresponds to case (b) in Fig. \ref{fig:schematicprefer}. Using equation (\ref{eq:MchiLimited}) and the minimum ratio between the SuperCDMS exclusion and the lower boundary of the CDMS II Si region, $\delta = 0.22$, we conclude that SuperCDMS excludes the CDMS II Si preferred region for all DM halos whenever $m_\chi \gtrsim 11$~GeV. This corresponds to the dot-dashed red line in \Fig{fig:ratioprefer}, left. 

Since $M_{{\rm Ge}} > M_{{\rm Si}}$, we can applying the escape velocity constraints from \Sec{sec:escape}. Using $p_{B,max} \simeq 2.1 \times 10^{-2} \ \GeV/c$, we find that SuperCDMS excludes CDMS II Si for all DM masses and all halos with escape velocity less than $v_{esc} \simeq 410 \ {\rm km}/{\rm s}$. Indeed, this can be seen in the left panel of \Fig{fig:vMinContour}: for $m_{\chi} = 10 \ \GeV$, the lower boundary of the oreferred envelope contains points with $v_{min} > 410 \  {\rm km}/{\rm s}$. Consequently, while the SuperCDMS exclusion does not cross the lower boundary of the envelope, a DM interpretation is still excluded for halos with sufficiently small escape velocities, since going to smaller DM masses would push the envelope further to the right.  This implies that a consistent interpretation even for $m_\chi \lesssim 11$~GeV will require scattering from DM particles in the tail of the DM velocity distribution above $410 \  {\rm km}/{\rm s}$. Of course, the Galactic halo almost certainly has a larger escape velocity than $410 \  {\rm km}/{\rm s}$, but we regard this example as an interesting proof-of-principle for a method which helps point towards particular DM mass ranges and halo properties.

\section{Analysis Recipe}
\label{sec:recipe}
In order to develop a halo-independent momentum-space analysis framework, we have had to introduce a good deal of notation, as well as several types of plots which were useful for pedagogical purposes. However, in practice the momentum-space method is straightforward, and here we summarize the approach. The minimal analysis recipe is as follows:
\begin{enumerate}
\item For each experiment, compute limits (for null results) or best-fit regions (for signals) in $p_R-\tilde{h}$ space as in \Fig{fig:pRContour}, left panel. This is accomplished by constraining the differential event rate in \Eq{eq:rateanew} using whichever statistical technique is desired. For each individual experiment, this represents the universal halo-independent momentum-space result which can be derived from the data and this plot is the same for all DM masses, i.e. it is mass-independent.\footnote{By definition this contains all of the information required to make the standard $v_{min}$-space plot for any specific choice of DM mass if desired.}
\item To compare the consistency of two experiments, construct the $R(\tilde{h})$ plot from the ratio of the $\tilde{h} (p_R)$ curves for each experiment as in \Fig{fig:pRContour}, right panel, using \Eq{eq:RPref}.  The value of $\delta$ (defined in \Eq{eq:deltadef}) extracted from such a plot determines the endpoint of the range of masses for which the two experiments are inconsistent, using \Eq{eq:MAB}.
\item As an optional extension, if the target mass for the null experiment is greater than the target mass for the experiment which observes a signal, determine the value of $v_{esc}$ for which the two experiments are inconsistent for all halos and all masses, using \Eq{eq:vescCrit}. The value of $p_{B,max}$ required can be read off the momentum-space plot constructed in step 1.
\end{enumerate}
Note that in the recipe we have described it is never necessary to pick a fiducial $m_\chi$ and actually construct the mapping into $\vmin$-space. In particular, the plots in \Fig{fig:vMinContour} are not necessary for the analysis, and were provided here only for pedagogical reasons. The momentum-space plot, which is unique and independent of the DM mass, contains all of the information necessary to construct the halo-independent $v_{min}$-space plot for any DM mass.  Thus, in the presentation of new experimental data, a single momentum-space plot contains all of the information required for external groups to perform halo-independent analyses.  In this way one automatically avoids introducing the errors in interpretation which may arise from those external to an experimental collaboration attempting to recreate quantitative details of experimental analyses.

\section{Conclusions}
\label{sec:conclusion}

In this paper we have shown how a simple change of variables, from $\vmin$-space to recoil momentum space, allows a presentation of halo-independent DM direct detection results without a fiducial choice of the DM mass. While previous applications of halo-independent methods required a choice for $m_\chi$, resulting in a proliferation of plots, the positive or null results of an experiment for all DM masses can be shown on a single $p_R - \tilde{h}$ plot. To compare results between experiments, an accompanying $R(\tilde{h})$ plot which shows the ratios of the various experimental constraints may also be useful. Such a plot is easily derived from the $p_R - \tilde{h}$ plot, although we emphasize that it this additional plot is not necessary for the analysis of a single experiment.  Depending on the mass orderings of the detector materials, one can in some cases determine directly from single $p_R - \tilde{h}$ plot, or in all cases from the $R(\tilde{h})$ plot, whether a DM interpretation for any possible halo is ruled out either (a) for all DM masses, or (b) for a limited range of DM masses, where the upper or lower endpoint of this range can be computed from an equation like \Eq{eq:MchiLimited} and read directly off the $R(\tilde{h})$ plot. 

In our discussion, only certain simplified scenarios have been sketched, and more complex exclusion curves and preferred regions are possible in the momentum-space plot. However, the generalization of our results is completely straightforward, and our sample analysis in \Sec{sec:example} (and \App{app:limits}) shows that our method can be easily applied to real data. As the momentum-space plot is unique, in that it does not need to be recreated for multiple DM mass values, we believe that it is ideally suited for any halo-independent analysis by an experimental collaboration. This would allow the presentation of experimental results in a way which is as complementary as possible to the $m-\sigma$ plots already in common use.

\acknowledgments{YK thanks Andrew Brown for fruitful discussions.  We thank Felix Kahlhoefer for very insightful and helpful comments on a earlier version of this work. This work is partially supported by the U.S.\ Department of Energy under cooperative research agreement Contract Number DE-SC00012567. YK is also supported by an NSF Graduate Fellowship. AJA is supported by a Department of Energy Office of Science Graduate Fellowship Program (DOE SCGF), made possible in part by  the American Recovery and Reinvestment Act of 2009, administered by ORISE-ORAU under contract no. DE-AC05-06OR23100.  Fermilab is operated by Fermi Research Alliance, LLC under Contract No. DE-AC02-07CH11359 with the United States Department of Energy.}

\appendix

\section{Comparisons Between Null Results}
\label{app:limits}
While the ultimate purpose of the mass-independent method is as a diagnostic tool in comparing emerging hints of DM detection with null direct detection results, in this appendix we consider the comparison between momentum-space exclusion curves for two different experiments.

\subsection{Ratio Plots for Null Results}
\label{sec:excludecomp}

We consider two direct detection experiments $A$ and $B$, with target nuclei of mass $M_A$ and $M_B$ respectively.  Since we are comparing two null results we can assume, without loss of generality, that $M_A>M_B$.\footnote{Note that this was not the case in comparing a null result to a putative positive result.}

\begin{figure}[t]
  \centering
 \includegraphics[height=0.37\textwidth]{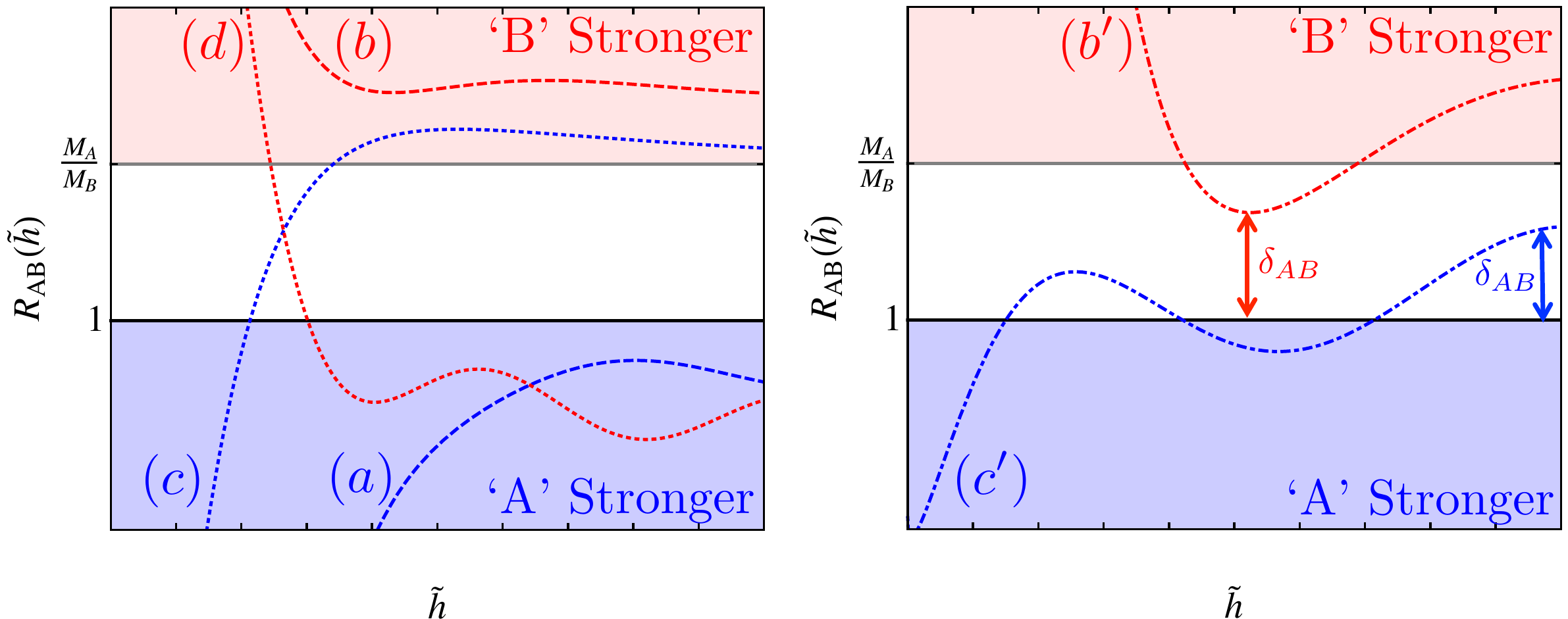} 
\caption{Some possibilities for the variable $R_{AB}$ defined in \Eq{eq:RAB}.
  After rescaling to $\vmin$-space, a point satisfying $R_{AB,\chi} <1$ ($R_{AB,\chi} >1$) signifies a stronger constraint from experiment $A$ ($B$).  This rescaling is given by $F_\chi$, defined in \Eq{eq:rat}, which satisfies the inequality \Eq{eq:ineq}.  Thus any curve lying entirely below $R_{AB} =1$, such as curve (a), implies that for all DM masses the halo-independent $v_{min}$-space constraint will be stronger for experiment $A$ than experiment $B$.  The converse is true for any curve that lies entirely above $R_{AB} =M_A/M_B$, such as curve (b).  See text for a detailed discussion of the other curves and general cases.}
  \label{fig:ratio}
\end{figure}

As discussed in \Sec{sec:prefercomp}, in some simple cases it is possible to determine the relative strengths of bounds from two different detectors directly from their $\tilde{h} (p_R)$ plots.  For instance, in \Fig{fig:mapping} experiment $A$ finds the strongest constraints in $p_R$-space, and as any mapping to $v_{min}$-space will shift the curve for $B$ further to the right relative to $A$, we may conclude that experiment $A$ finds the strongest exclusion for all DM masses.  However, in general the two exclusion curves may in principle cross any number of times while remaining consistent with the monotonically decreasing behavior. 

To treat the general case we again exploit the fact that mapping from momentum-space to $v_{min}$-space only shifts curves relative to each other in the horizontal direction.  Each momentum-space exclusion curve is given by the function $\tilde{h} (p_R)$; since this curve is monotonically decreasing, we may again define the inverse mapping $p_R (\tilde{h})$ as in \Eq{eq:hmap}. We now define the ratio of the momentum-space exclusions for the two experiments
\be
R_{AB}(\tilde{h}) = \frac{p_{R,A} (\tilde{h})}{p_{R,B} (\tilde{h})}  ~~,
\label{eq:RAB}
\ee
which is the analogue of $R(\tilde{h})$ defined in \Eq{eq:RPref}. The subscript $AB$ is meant to indicate the mass ordering, $M_A > M_B$.

\begin{figure}[t]
\centering
{\includegraphics[width=3.0in]{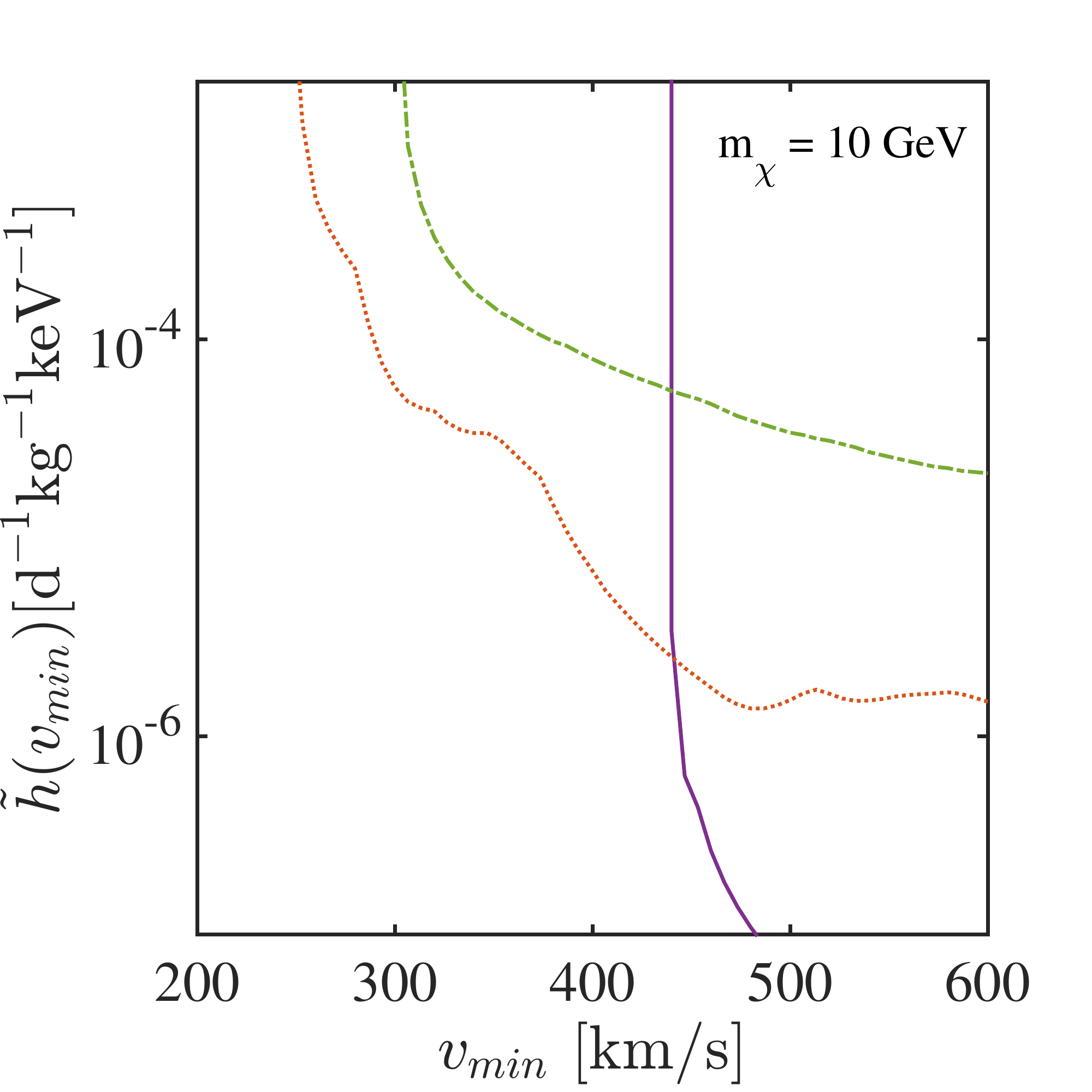}}
{\includegraphics[width=3.0in]{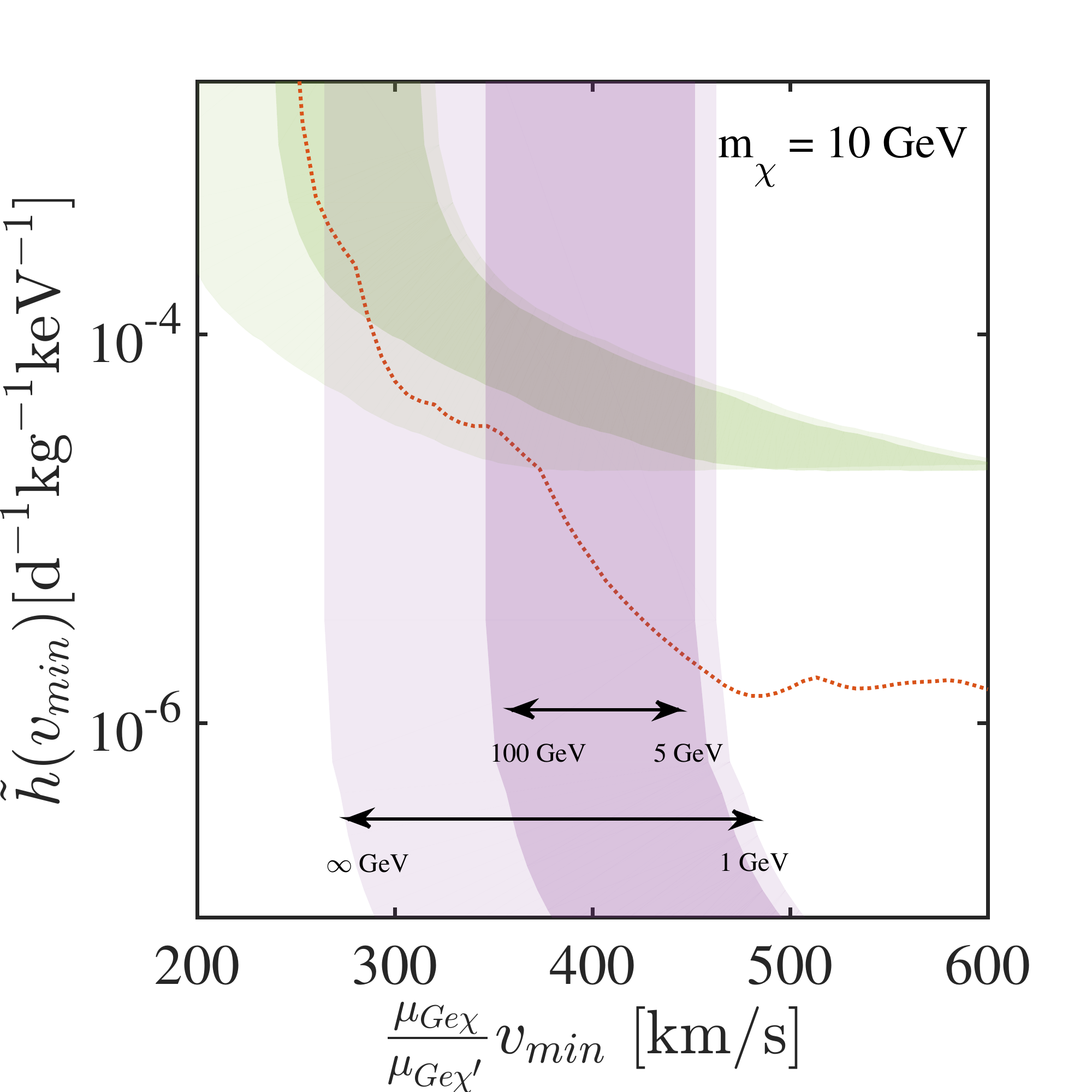}}
\caption{Exclusion limits from XENON10 (dot-dashed green), LUX (solid purple), and SuperCDMS (dotted orange) in $v_{min}$-space. {\bf Left}: $v_{min}$-space exclusions for 10~GeV DM. Kinks and slight deviations from monotonicity are present in the SuperCDMS limit, occuring when the ``Pmax'' method switches the interval used to calculate the exclusion. Similar discrete artifacts are not present in LUX because an observation of zero events is used to calculate the limit, nor in XENON10 because the experiment observed a large number of events. {\bf Right}: limits with $v_{min}$ calculated for a range of DM masses and scaled to $m_\chi = 10$~GeV using the factor appropriate for Ge $\mu_{Ge,\chi} / \mu_{Ge,\chi^\prime}$, where $\mu_{Ge,\chi}$ is the reduced mass for 10~GeV DM and Ge target. This has the effect of mapping SuperCDMS limits to a single line (dotted orange), independent of mass, while limits from Xe experiments are mapped to different locations depending on the DM mass. Dark bands correspond to the range of limits with masses 5-100~GeV, while the light bands show the range of limits for all possible masses, which facilitates comparison of the relative strengths of the limits for different DM masses.}
\label{fig:vMinLimits}
\end{figure}

It is useful to consider first the general form of the $R_{AB}$ curves.  In general, two experimental exclusion lines in momentum-space may cross any number of times, and thus $R_{AB}$ may cross unity and change gradient any number of times. In particular, $R_{AB}(\tilde{h})$ need not be monotonic. Its asymptotic behavior is somewhat simpler, though. As $\tilde{h} \to \infty$, the asymptotic value of $R_{AB}(\tilde{h})$ is the ratio of the low-energy thresholds of experiments $A$ and $B$.  The behavior in the other extreme $\tilde{h}\rightarrow 0$ is determined by the constraint which extends to lower $\tilde{h}$ at high $p_R$; if $A$'s constraint extends to lower $\tilde{h}$, then $R_{AB}(\tilde{h}\rightarrow 0)\rightarrow 0$, otherwise $R_{AB}(\tilde{h}\rightarrow 0)\rightarrow \infty$. These two cases inform the definition of $\delta_{AB}$, the analogue of $\delta$ defined in \Eq{eq:deltadef}: 
\be
\delta_{AB} =  \left\{ \begin{array}{lr}
R_{AB}^{{\rm min}}(\tilde{h}) - 1, \ {\rm if\ } R_{AB}(\tilde{h}\rightarrow 0)\rightarrow \infty \\
R_{AB}^{{\rm max}}(\tilde{h}) - 1, \  {\rm if\ } R_{AB}(\tilde{h}\rightarrow 0)\rightarrow 0~~.
\end{array}
\right.
\ee
Here, $\delta_{AB}$ is always positive since we have assumed $M_A > M_B$. For curves that have $R_{AB}(\tilde{h}\rightarrow 0)\rightarrow 0$, $\delta_{AB}$ is the furthest distance the curve gets above 1, and for $R_{AB}(\tilde{h}\rightarrow 0)\rightarrow \infty$, $\delta_{AB}$ is the distance of closest approach to 1. 

With these general discussions in hand we now discuss the various classes of behaviour for the $R_{AB}$ curves, illustrated in \Fig{fig:ratio}. As in \Sec{sec:prefercomp}, the rescaling of $R_{AB}$ into $\vmin$-space is $R_{AB,\chi} = R_{AB} F_\chi$, with $F_\chi = \mu_{B\chi}/\mu_{A\chi}$ satisfying $M_B/M_A \leq F_\chi \leq 1$. We first discuss the cases in \Fig{fig:ratio}, left:
\begin{enumerate}
\item [\textbf{(a)}]  
Since $R_{AB}<1$ everywhere then for all DM masses $A$ would give the strongest constraint over all of $v_{min}$-space, as $B$ would shift further to the right than $A$ in all cases.  This is the case shown in \Fig{fig:mapping}. $A$ gives the strongest constraint over all of $v_{min}$-space, and thus for all possible DM halos for all DM masses.
\item [\textbf{(b)}]  
This is the complementary case to (a) as $R_{AB}>M_A/M_B$ everywhere.  The rescaling in going to $\vmin$-space is insufficient to alter the relative strength of the constraints, for all DM masses.  In going to $v_{min}$-space, the smallest factor by which the curve (b) is rescaled is $M_B/M_A$, corresponding to $M_\chi \to \infty$.  This cannot push any point on (b) below unity, so in this case $B$ gives the strongest constraint over all of $v_{min}$-space and hence for all possible DM halos and all DM masses.
\item [\textbf{(c)}]  
This case corresponds to a situation where, for all DM masses, the two experiments provide complementary constraints.  There is no DM for which either experiment will give the strongest constraints over all of $v_{min}$-space: at low $\tilde{h}$, $A$ is a stronger bound, and at high $\tilde{h}$, $B$ is a stronger bound.  
\item [\textbf{(d)}]  
The same as (c) but with the roles of $A$ and $B$ reversed: at low $\tilde{h}$, $B$ is a stronger bound, and at high $\tilde{h}$, $A$ is a stronger bound.  

\end{enumerate}

Cases (b$^\prime$) and (c$^\prime$) in \Fig{fig:ratio}, right, are those for which $A$ or $B$ can give the strongest constraint for a range of DM masses. As in \Sec{sec:prefercomp}, the endpoints of these ranges are determined by $M_{AB}(\delta_{AB})$, defined in \Eq{eq:MAB}.
\begin{enumerate}
\item [\textbf{(b$^\prime$)}]  
If the DM mass mass satisfies 
\be\label{eq:caseb}
M_\chi <  M_{AB}(\delta_{AB})  ~~,
\ee
$B$ will give the strongest constraint over all of $v_{min}$-space and hence for all DM halos.
\item [\textbf{(c$^\prime$)}]  
If the DM mass satisfies
\be
M_\chi > M_{AB}(\delta_{AB})  ~~,
\ee
then $A$ will give the strongest constraint over all of $v_{min}$-space and hence for all DM halos.
\end{enumerate}

\subsection{A Sample Analysis: Comparing LUX, XENON10, and SuperCDMS}
The dark matter mass scaling relations introduced in section \ref{sec:excludecomp} for comparing null results can be used to concisely illustrate the mass ranges for which Xe-based and Ge-based experiments give stronger results. In the left panel of Fig. \ref{fig:vMinLimits} we compare limits from XENON10, LUX, and SuperCDMS, for a 10~GeV DM. As was demonstrated in \Sec{sec:example}, this can be visualized in a single plot, such as the right panel of Fig. \ref{fig:vMinLimits}, which shows the range of limits for each experiment, when referred to 10~GeV DM using the scaling factor appropriate for Ge. Such a plot will contain, for each possible mass, the correct positioning of LUX or XENON10 limits relative to SuperCDMS. Although one cannot directly infer the value of $v_{min}$ at which curves cross at each DM mass, it is manifestly clear from the plot that SuperCDMS is always stronger than XENON10 at small DM masses, and that SuperCDMS and LUX are complementary across all DM masses due to the higher $v_{min}$ threshold of LUX.

\begin{figure}[t]
  \centering
\includegraphics[width=3.0in]{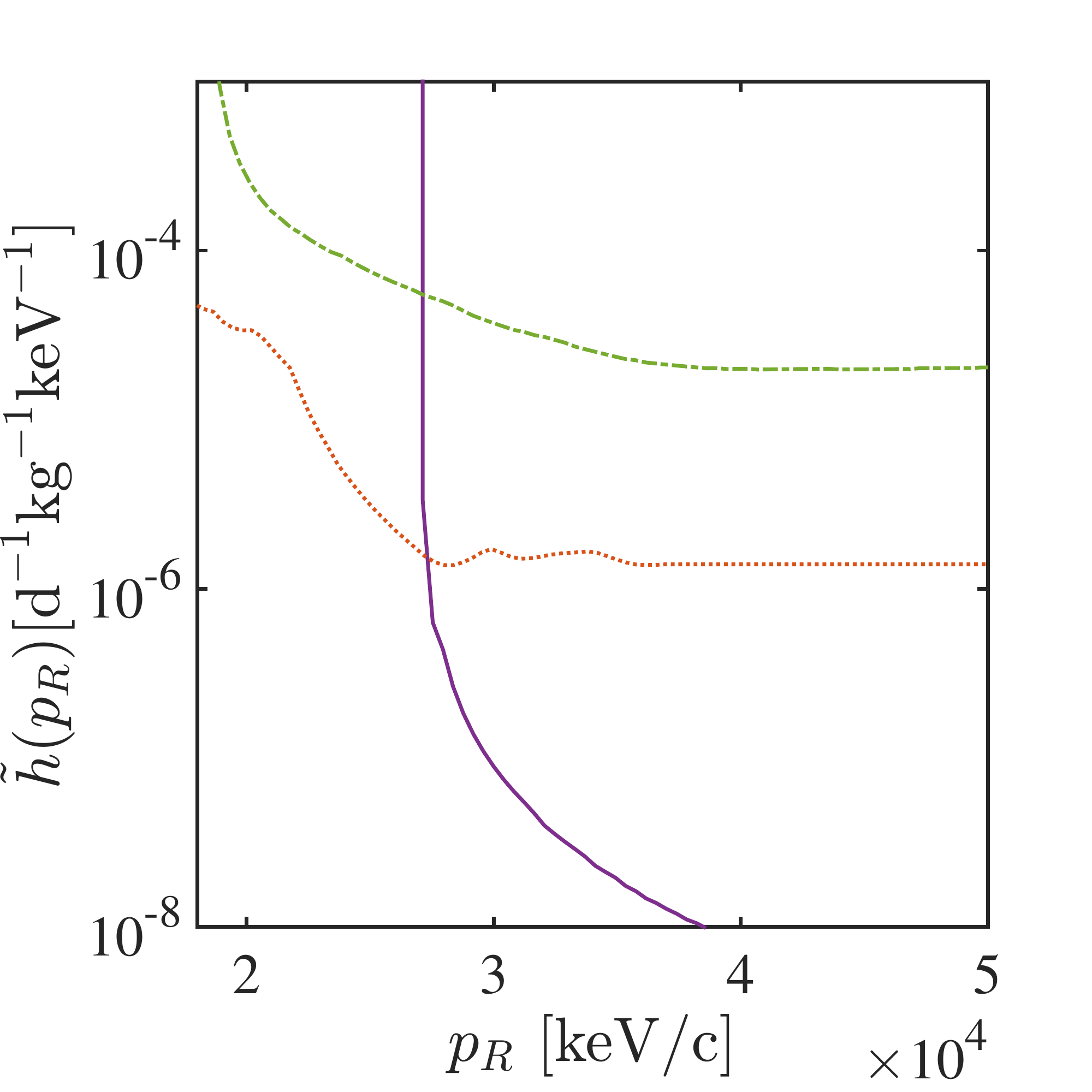}
\includegraphics[width=3.0in]{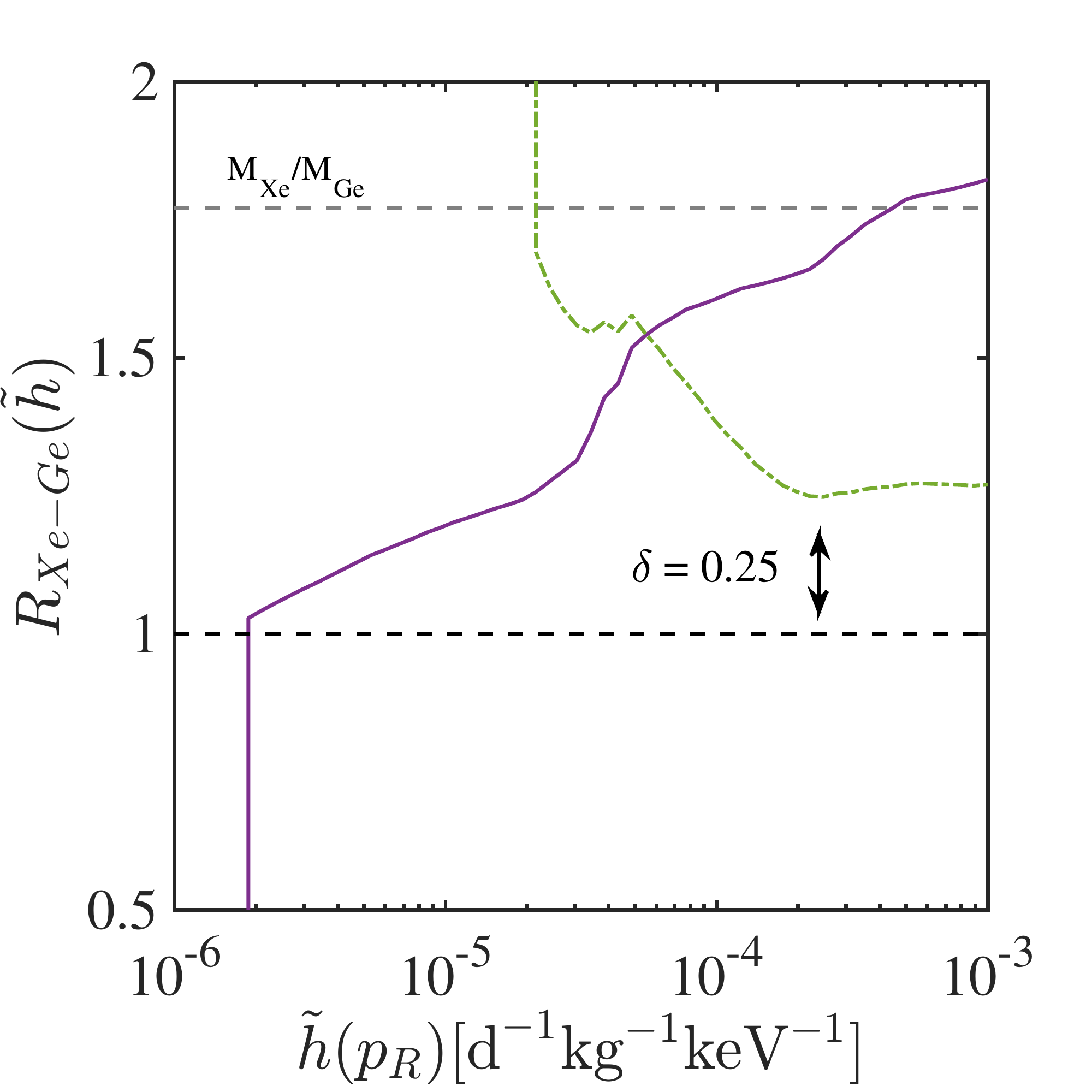}
\caption{Comparison of mass-independent exclusion limits. {\bf Left}: momentum space exclusion limits from XENON10 (dot-dashed green), LUX (solid purple), and SuperCDMS (dotted orange). {\bf Right}: ratio of exclusion in $p_R$-space of LUX to SuperCDMS (solid purple) and XENON10 to SuperCDMS (dot-dashed green). Deviations from monotonicity are smoothed so that the limits $\tilde{h}(p_R)$ can be inverted to form $R_{{\rm Xe-Ge}}(\tilde{h})$, as in equation (\ref{eq:RAB}). The minimum value of $\delta = 0.25$ implies that SuperCDMS is stronger than XENON10 across the full range of $v_{min}$ only for $m_\chi \lesssim 59$~GeV. The SuperCDMS and LUX limits, on the other hand, probe complementary ranges of $v_{min}$ for all DM masses.}
\label{fig:pRLimits}
\end{figure}

Fig. \ref{fig:pRLimits} shows the momentum-space exclusion plots for SuperCDMS, XENON10, and LUX. The left panel shows the exclusions as a function of recoil momentum, while the right panel shows the ratio of LUX and XENON10 limits to the SuperCDMS limit, as a function of $\tilde{h}$. Comparing Fig. \ref{fig:pRLimits} to the cases enumerated in Fig. \ref{fig:ratio} indicates that the pair of limits from SuperCDMS and LUX corresponds to case (c), while the pair of SuperCDMS and XENON10 corresponds to case (b$^\prime$). In case (c), both limits are complementary in $v_{min}$ space for all possible DM masses, and this is clearly apparent for SuperCDMS and LUX in Fig. \ref{fig:vMinLimits}. In case (b$^\prime$), SuperCDMS is stronger than XENON10 for DM masses below the bound in equation (\ref{eq:caseb}). Using the value of $\delta = 0.25$ read off from the $R_{{\rm Xe-Ge}}$ plot constructed from the SuperCDMS and XENON10 limits, we find that SuperCDMS is stronger for all $v_{min}$ when $m_\chi \lesssim 59$~GeV. This behavior is also qualitatively visible in the right panel of Fig. \ref{fig:vMinLimits}.

\bibliographystyle{JHEP}
\bibliography{massind}{}

\end{document}